\documentclass[preprint,10pt]{elsarticle}

\usepackage{amssymb}
\usepackage{amsfonts}
\usepackage{epsfig}
\usepackage{helvet}
\usepackage{courier}
\usepackage{amssymb}
\newtheorem{theorem}{Theorem}[section]

\journal{Communications in Nonlinear Science and Numerical Simulation}

\begin{document}

\begin{frontmatter}

\title{Heavy-tailed Distributions In Stochastic Dynamical Models}
\author{Ph. Blanchard${}^{1,2}$, T. Kr\"{u}ger${}^{2}$
D. Volchenkov${}^{2}$
\footnote{E-Mail:{\it  volchenk@physik.uni-bielefeld.de}}
}
\maketitle

\begin{enumerate}
\item[${}^1$]{\it Department of Physics, Universit\"{a}t
 Bielefeld, Postfach 10 01 31, 33501 Bielefeld, Germany}
\item[${}^2$]{\it Center of Excellence Cognitive Interaction Technology, Universit\"{a}t Bielefeld, Postfach 10 01 31, 33501 Bielefeld, Germany }
\end{enumerate}

\begin{abstract}

Heavy-tailed distributions
 are found throughout many naturally
 occurring phenomena.
We have reviewed
 the models of stochastic dynamics  
that lead to heavy-tailed distributions
(and power law distributions, in particular)
including   
the multiplicative noise models,
the models subjected to the
Degree-Mass-Action principle (the generalized 
preferential attachment principle),
 the intermittent behavior occurring in 
 complex physical systems near a bifurcation point, 
 queuing systems, 
and 
the models of Self-organized criticality.
Heavy-tailed  distributions
appear in them as the 
 emergent phenomena
sensitive for 
coupling rules essential for
 the entire 
dynamics. 
\end{abstract}

\vspace{0.5cm}


\end{frontmatter}

\tableofcontents

\section{Introduction}
\label{sec:Introduction}
\noindent

In 1906,
a lecturer in economics
 at the University of Lausanne in Switzerland
V.F. Pareto
had discovered that
 the allocation of wealth among individuals can be 
efficiently described by a  {\it power law} probability distribution,
 as being self-similar over a wide range of wealth magnitudes.
A great many 
such distributions have been found in diverse
fields of science being 
a long-known feature of many
empirical distributions 
such as Pareto, Zipf, and L\'{e}vy
distributions used to model real-world phenomena.
In particular, 
 Paul L\'{e}vy 
worked on a class 
of probability distributions with "heavy tails",
 which he called
{\it stable distributions}
 largely considered
probabilistic curiosities 
at the time, as
heavy-tailed distributions
 have properties that are qualitatively different
 to the many commonly used
distributions such as
exponential, normal or Poisson.

\begin{figure}[ht!]
 \noindent
\begin{center}
\epsfig{file=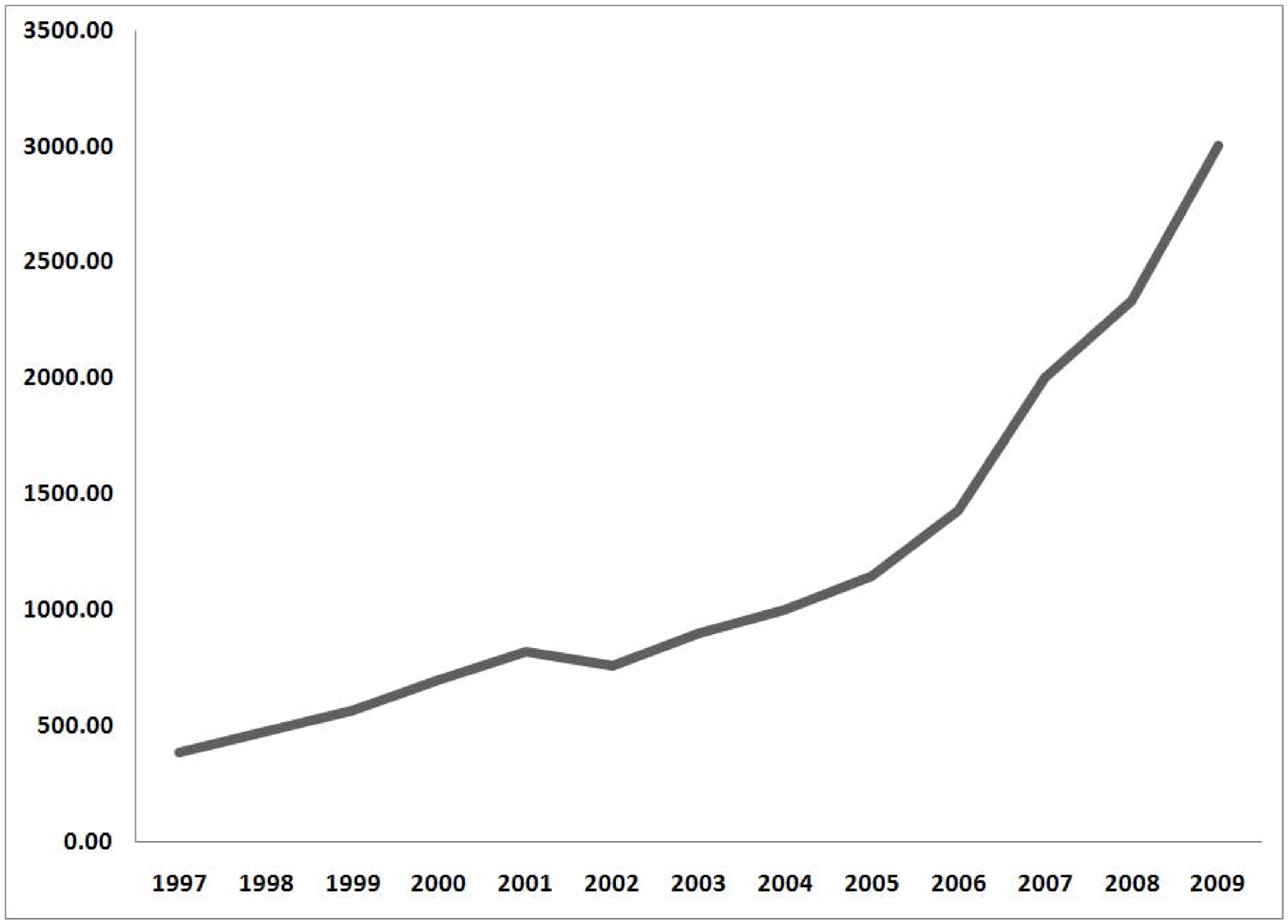,  angle= 0,width =8cm, height =5.5cm}
  \end{center}
\caption{\footnotesize 
The total number of publications 
devoted to the 
heavy-tailed
and power law
distributions  in the 
arXiv ({\it http://arxiv.org}) grows fast each year.
\label{Fig_trend}}
\end{figure}

Since then, there has 
 been a permanent surge of interest to heavy-tailed
and power law
distributions  
 from scientists
working in  fields as diverse as
weather forecasting
to stock market analysis.
The growth rate  
of the
interest to 
the 
heavy-tailed
and power law
distributions  
    can be attested by 
the yearly increase 
of the
total number of publications 
on the topic (see Fig.~\ref{Fig_trend})
in the 
arXiv ({\it http://arxiv.org}),
the major forum for
disseminating scientific results in Physics, Mathematics, Nonlinear
Sciences, Computer Science, and Quantitative Biology.

The purpose of this review paper is to explain 
 the models of stochastic dynamics  
that lead to heavy-tailed and power law distributions.  

In this survey, we
call a distribution  
{\it heavy-tailed}
if it does not have the exponentially bounded
asymptotes.
Namely, 
given a random variable $X$
characterized by the probability
distribution function
$$
F(x)\,\,=\,\,\Pr\left[x<X\right],
$$
we say that it has a {\it heavy right tail} if
\begin{equation}
\label{fat_right_tail}
\lim_{x\to\infty} F(x)\cdot\exp\,\lambda x\,\,=\,\,\infty,\quad \forall \lambda>0.
\end{equation}
An important subclass of heavy-tailed distributions 
is the {\it sub-exponential} distributions
\cite{Embrechts:1997}, for which 
 a sum of $n$ independent random variables
$X_1,$ $X_2,$ $\ldots,X_n,$
  with common distribution $F(x)$
is given,
\begin{equation}
\label{subexp}
\Pr\left[x< X_1+X_2+\ldots+X_n\right]\,\,\begin{array}{c}
\sim \\ {}_{x\to \infty}
\end{array}\,\,
\Pr\left[ x< \max\left(X_1,X_2,\ldots,X_n\right)\right].
\end{equation}
Among sub-exponential distributions, 
we shall be essentially interested in 
those  distributions 
of a random variable $X$ 
which 
are characterized by a {\it power law decay},
\begin{equation}
\label{fat-tailed}
\Pr\left[x<X\right]\,\,\begin{array}{c}
\sim \\ {}_{x\to \infty}
\end{array}\,\,\frac{1}{\,\,x^{1+\gamma}\,\,},
\end{equation}
for some $\gamma>0.$

There are several physical mechanisms
 known as underlying the 
power law behaviors. 
Power law distributions 
 often manifest a form of regularity
arising through growth processes
 which is composed of a large number
 of common events 
and a small number of rarer events happened randomly.
As the number of events grows,
their distribution,
 under certain conditions,
might converge to a steady state
(\ref{fat-tailed}).
The mechanism of {\it preferential attachment}
 in which some quantity is distributed
at random among a number of individuals 
according to how much they already have had been proposed in
 \cite{Barabasi:1999}
  as an algorithm explaining power law degree
 distributions in some scale-free networks.
Graphs in that 
 grow by successively adding a new vertex
say $x$ at each time step that links to an
already existing one of the degree $k$,
 with the probability 
 $\propto k/N$
where $N$ is the total number of vertices present in the graph.
The preferential attachment process generates
 a heavy-tailed distribution following a  power law in its tail. 

Power laws are also found in the study 
of stochastic processes involving
multiplicative noises \cite{Levy:1996}.
A typical equation of multiplicative stochastic
process is given by a linear {\it Langevin equation}
with a randomly changing coefficient. 
The effect of such
a random coefficient
drastically enhance
the additive random force in the Langevin equation
and increase fluctuations.
Following \cite{Takayasu:1997},
 let us consider temporal fluctuations for 
a simple discrete time version of
the Langevin equation,
\begin{equation}
\label{discrete_langevin}
x(t+1)\,\,=\,\, b(t)\cdot x(t) +f(t),
\end{equation}
where $f(t)$ represents a random additive noise,
and $b(t)$ is a non-negative stochastic coefficient 
interpreted as
dissipation for $b(t)<1$ and amplification
 for $b(t)>1.$
  If we assume 
for simplicity that $b(t)$ and $f(t)$
are independent white noises having stationary statistics,
and $f(t)$ is symmetric,
we obtain,
 for the second order moment,
\begin{equation}
\label{second_order_langevin}
\left\langle
x^2(t+1)\right\rangle\,\,=\,\,
\left\langle b^2\right\rangle
\left\langle x^2(t)\right\rangle\,+\,
\left\langle f^2\right\rangle,
\end{equation}
in which the angular brackets denote an average over realizations.
For constant $b^2$ and $f^2$, there is a stationary solution,
\begin{equation}
\label{stationary_langevin}
\left\langle x^2\right\rangle\,\,=\,\,
\frac{\left\langle f^2\right\rangle}
{\,\,1- \left\langle b^2\right\rangle\,\,},
\quad \left\langle b^2\right\rangle<1,
\end{equation}
but $\left\langle x^2(t)\right\rangle$ diverges as 
$t\to\infty,$ for $\left\langle b^2\right\rangle>1.$
The statistics of $x(t)$ is estimated theoretically
by introducing the characteristic function,
\begin{equation}
\label{char_fun_langevin}
\mathfrak{F}(\xi,t)\,\,=\,\,\left\langle
e^{i\xi\, x(t)}
\right\rangle.
\end{equation}
When 
$\left\langle x^2(t)\right\rangle$ diverges,
$\mathfrak{F}(\xi,t)$ has singularity at
 $\xi= 0$ in the limit of $t\to \infty,$
so that Taylor expansion cannot be applied for the steady
solution. In such a case the following fractional power
term can be assumed for the lowest order term because
the characteristic function is generally a continuous function
\begin{equation}
\label{taylor_char_fun_langevin}
\mathfrak{F}(\xi,\infty)\,\,=\,\, 1- \mathrm{const}\cdot \left|\xi\right|^\beta+\ldots, \quad 0<\beta<2,
\end{equation}
which is equivalent to the assumption of power law tails
for the cumulative probability distribution
\begin{equation}
\label{cummulative-langevin}
\Pr\left[\,|x|<X\,\right]\,\,\sim\,\,\frac{1}{\,\,x^\beta\,\,}.
\end{equation}
The validation of stochastic mechanisms generating the power-law
behavior remains an active field of research in many areas of
modern science. 

In the forthcoming sections, we 
consider in details 
other, more sophisticated mechanisms
that might bring forth  
 fat--tail statistics
 in dynamical systems.

The plan of our review is following. 
In Sec.~\ref{sec:FAT_TAILS_1}, 
we discuss the 
Degree-Mass-Action principle in random graphs formation - 
the preferential attachments and their natural 
generalizations. 
In Sec.~\ref{sec:FAT_TAILS_2}, 
we consider 
the statistics of bursts in 
systems close to a threshold of instability. 
Then, in Sec.~\ref{sec:FAT_TAILS_3},
we review the emergence of fat tails in queuing systems.
Finally, in Sec.~\ref{sec:FAT_TAILS_4}, 
we explain the appearance of
power law distributions in
models of Self--Organized Criticality.
We conclude in the last section. 

\section{Degree-Mass-Action principle in random graphs formation }
\label{sec:FAT_TAILS_1}
\noindent

Random graphs with  scale--free probability
 degree distributions are ubiquitous while modeling 
many real world networks such as 
the
World Wide Web, social, linguistic, 
citation and biochemical networks;
 an excellent  
 survey is  \cite{Albert:2002}.
The preferential attachment principle,
  together with its various 
 modifications,
could be seen as a particular case of 
{\it degree-mass-action}
 principle since
 the degree of a node acts in that 
 as a  
positive affinity parameter
 quantifying the attractiveness of the node for new vertices.
Our first aim
 is to construct a family
 of static
random graph models,
 in which vertex degrees 
are distributed power-law like,
while edges still have high
 degree of independence. 
As usual in random graph
theory,
 we will entirely deal 
with asymptotic properties
 in the sense that
the graph size goes to infinity.

\subsection{A random graph space of the preferential attachment model \label{subsec:A_random_graph_space_of_the_preferential-choice_model}}
\noindent

We consider graphs
 with vertex set $V=V_{n}$ $=\left\{ 1,...,n\right\} $
where an edge between the vertices $x$ and $y$ 
(denoted by $x\sim y$) is
interpreted as a persistent 
contact between the two nodes.
 Given $x\in V$,
its degree is denoted by $d(x)$.

 We will think of edges as generated by a
pair-formation process in which each vertex $x$ - 
often denoted as an
individual - chooses a set of partners according
 to a specified 
$x$-dependent rule. 
Therefore the set of individuals
 which have contact with a
given vertex $x$ can be divided into
 two -possible non-disjoint sets: 
the
set of nodes which are chosen by $x$ himself 
and the set of nodes which have
chosen $x$ as one of their partners. 
We call the size of the first set the
out-degree $d_{out}(x)$ of $x$  
and the size of the second one the in-degree 
$d_{in}(x)$ of $x$. 
Obviously, 
\begin{equation}
\label{deg_deg_in_deg_out}
d(x)\,\,\leq\,\, d_{out}(x)+d_{in}(x),
\end{equation}
 and if the
choices are sufficiently independent one can 
expect equality to hold almost
surely if $ n\rightarrow \infty $.

We partition the set of vertices $V_{n}$
 into groups $\left\{
C_{i}(n)\right\} _{i\geq 1}$ where all members
 of a group $C_{i}(n)$ choose
exactly $i$ partners by themselves
 ($d_{out}=i$ on $C_{i}(n)$). 
Let $
P_{\alpha }^{1}\left( n,j\right) $ 
the probability for $x$ to choose a fixed
partner $y\in $ $C_{j}\left( n\right) $
 if $n$ partners are available for
the choice and just one choice will be made be 
\begin{equation}
P_{\alpha }^{1}\left( n,j\right) \;=\;A_{\alpha }(n)\mathbb{\ }\frac{%
j^{\alpha }}{n}
\end{equation}
Here $A_{\alpha }(n)$ is a normalization constant such that 
$$
A_{\alpha
}(n)\left( \sum_{i\geq 1}\;|C_{i}(n)|\frac{i^{\alpha }}{n}\right) =1
$$ 
and $\alpha $ is a real parameter. Since we want 
$$
A_{\alpha }(n)\rightarrow_{n\to \infty}
A_{\alpha }
$$ 
we need 
$\sum_{i}|C_{i}(n)|\frac{i^{\alpha }}{n}$
 to be bounded as a function of $n$ which will impose later
on constraints on the constant $\alpha $.
The parameter $\alpha $ acts as an
affinity parameter tuning the tendency to
 choose a partner with a high
out-degree or low out-degree. If $\alpha =0$, 
choices are made without any
preferences and $A_{\alpha }(n)\equiv 1$.
 For $\alpha >0$ the "highly
active" individuals are preferred whereas 
for $\alpha <0$ the "low
activity" individuals are favored.
 From this we obtain the basic
probability 
\begin{equation}
\Pr [x\sim y ]\,\,\simeq\,\, A_{\alpha }(n)\,
\frac{i\cdot j^{\alpha }}{n}, \quad x \in C_{i},\quad
y \in C_{j}.
\end{equation}
Concerning the size of the sets $C_{i}(n)$ we will make the following
assumption: 
\begin{equation}
\frac{|C_{i}(n)|}{n}\,\,=\,\,p_{i}\left( n\right)\,\,\rightarrow _{n\rightarrow \infty }\,\,\frac{c_{1}}{\,\,i^{\gamma }\,\,}.
\end{equation}
With this choice we have to impose
 the restriction $\alpha <\gamma -1$ to
ensure the convergence of $A_{\alpha }\left( n\right) $. 
We require
furthermore $\gamma >2$ throughout the paper since otherwise the expected
in-degree for individuals from a fixed group would diverge. 
The basic
probabilities together with the fixed out-degree distribution define a
probability distribution on each graph with vertex-set $V_{n}$, and
therefore a random graph space 
$\mathcal{G}_{n}\left( \alpha ,\gamma \right)$.

First we want to compute the important pairing probabilities. 
We start with the easier case $\alpha =0$. 
\begin{equation}
\Pr \left( x\sim y\mid x\in C_{i};y\in C_{k}\right)\,\,
 =\,\,\frac{i+k}{n}- 
\frac{ik}{n^{2}}
\,\,{\sim}_{n\rightarrow \infty }\,\,\frac{i+k}{n 
} 
\end{equation}
Likewise one can compute the corresponding probabilities for $
\alpha \neq 0$. Dropping the simple details we just state the result:
\begin{equation}
\Pr \left( x\sim y\mid x\in C_{i}\ ;y\in C_{k}\right)\,\, 
\simeq\,\, A_{\alpha }\,\frac{
\left( ki^{\alpha }+k^{\alpha }i\right) }{n}, 
\quad n\rightarrow \infty .  \label{55}
\end{equation}
It turns out,that the typical graphs in this model 
still have for $\alpha <2$
a \emph{power-law }distribution for the degree 
with an exponent which can be
different from the exponent of the out-degree. 

For $\alpha >2$ we obtain a
degree distribution which follows in mean
 a power law but has gaps. To
compare both domains we will use the integrated tail distribution 
$$
F_{k}\,\,=\,\,\Pr \left( d\left( x\right) >k\right) .
$$ 
We will show that in both
cases we get the same integrated tail distribution.
 Since we are interested
in the dependence of the epidemic threshold from 
the power-law exponent of
the total degree distribution we have to analyze
 how this exponent varies
with the two parameters $\alpha $ and $\gamma  $
Since the partner choice
is sufficiently random and not too strongly biased
 toward high degree
individuals (that's the meaning of the condition
 $\alpha \leq \gamma -1$) it
is easy to see that the in-degree distribution of
 a vertex from group $C_{i}$
converges for $n\rightarrow \infty $ to a Poisson
 distribution with mean 
$\mathrm{const}\cdot i^{\alpha }$. 
There are essentially two regimes in the parameter
space, one for which the expected in-degree is of
 smaller order than the
out-degree over all groups and one where the in-degree 
is asymptotically of
larger order. In the first case it is clear that the
 in-degree is too small
to have an effect on the degree distribution exponent.
 In other words: the
set of individuals with degree $k$ consists mainly 
of individuals whose
out-degree is of order $k$. An easy estimation using
 the formula for the
pairing probabilities shows that the expected
 in-degree of individuals from
group $i$ is given by 
$$
\mathbb{E}\left( d_{in}\left( x\right) \mid x\in
C_{i}\right) \simeq \mathrm{const}\cdot i^{\alpha }
$$ 
asymptotically. Therefore the
in-degree is of smaller order than the 
out-degree if $\alpha <1$. In the case 
$\gamma -1\geq \alpha \geq 1$ the set of
 individuals with degree $k$ consists
mainly of individuals from groups with an 
index of order $k^{{1}/{\alpha}}$.

\subsection{The Cameo Principle. 
The origins of scale-free graphs in social
networks.  \label{subsec:CAMEO}}
\noindent

In the present section, 
we describe an edge formation principle
related upon 
a structure {\it apriori} imposed on the vertex set.
We  assume that  such a structure can be 
specified by 
a real positive random variable $\omega \in {\mathbb R}^{+}$
that quantifies 
some social property
of an individual such as
  its wealth, popularity, or 
beauty being distributed over the population 
with a given probability distribution 
$\varphi \left( \omega \right)$.

Furthermore, 
we assume that a link between the two individuals,
 $x$ and $y$, arises as a result of a directed choice
made by either $x$
or $y$ (symbolized by $x\rightarrow y$ or $y\rightarrow x$ respectively); 
in many real life networks edges are formed that
way.
Although 
the edge creation is certainly a directed process, in the present section
we consider
the resulting graph to be undirected  since for the majority of relevant
transmission processes defined on the network
 the original orientation of an edge is
irrelevant.

We  suppose that
the pairing probability follows an {\it inverse mass-action-principle}:
the probability that 
 $x$ decides to connect to $y$ characterized
by its affinity value $\omega \left( y\right) $ reads as 
\begin{equation}
\label{Cameo_probab}
\Pr \left\{ x\rightarrow y\mid \omega \left( y\right) \right\}\,\, \sim\,\,
\frac{1}{\,\,N\,\,}\cdot
\frac{1}{\,\,\varphi \left( \omega
\left( y\right) \right)^{ \alpha }\,\,},\quad \alpha \in \left( 0;1\right) 
\end{equation}
 where $N$ being the total number of vertices.
Let us note that it 
is not the actual value $\omega \left( y\right)$
which plays a
decisive role while pairing, but 
rather 
its relative frequency of appearance over the population.
The proposed principle 
captures the essence of antiquity markets  -- 
the more
rare a property is, the higher is its value, and the
 more attractive it becomes for others.
The paring probability model described above 
had been introduced in \cite{Blanchard:2004}
and called  
 the {\it Cameo-principle}
having in mind the
attractiveness, rareness and beauty of 
the small medallion with a profiled
head in relief called Cameo. 
And it is exactly their rareness and beauty
which gives them their high value.

In the  {\it Economics of Location theory} introduced by
\cite{Loesch:1954} and developed by \cite{Henderson:1974}, a city,
or even more certainly, a particular district in that
 may specialize in the production of a good 
that can be connected with natural resources, education,
policy, or just
low salary expenditures.
City districts compete among
themselves in a city market not necessarily
 connected with the quantity of their inhabitants.
The demand for these products
 comes into the city district from everywhere 
 and can be considered as exogenous.
Then, the   local
attractiveness of a site determining the creation of new spaces
of motion in that
 is
specified by a real positive random variable $\omega >0$.
Indeed, it is
rather difficult if ever be possible to estimate exactly
 the actual value $\omega(i)$ for any
 site in the urban pattern, since such an estimation can be referred to
both the economic an cultural factors
 that may
vary over the
different historical epochs and
 over the certain groups of population.
 In the framework of a probabilistic approach,
 it seems natural to consider the
value $\omega$ as a real positive
 independent  random variable
distributed over the
vertex set of the graph representation of the
site uniformly
 in accordance to
a smooth monotone decreasing probability
density function
$f \left( \omega \right)$.

While investigating the model of Cameo graphs, we 
assume that 

\begin{itemize}
\item  The parameter $\omega $ is independent identically distributed
(i.i.d.) over the vertex set with a smooth monotone decreasing density
function $\varphi \left( \omega \right) $

\item  Edges are formed by a sequence of {\em choices}. By a choice we
mean that a vertex $x$ chooses another vertex, say $y$ , to form an edge
between $y$ and $x.$ A vertex can make several choices. All choices are
assumed to be independent of each other.

\item  If $x$ makes a choice the probability of choosing $y$ depends only on
the relative density of $\omega \left( y\right) $ and is of the form (\ref{Cameo_probab}).

\item  A pre-defined out-degree distribution determines the number of choices
made by the vertices. The total number of choices (and therefore the number
of edges) is assumed to be about $\mathrm{const}\cdot $ $N.$
\end{itemize}

We focus on the striking observation that under the above assumptions
a scale-free degree distribution emerges
   independently of the particular choice of the $\omega-$
distribution.
Furthermore,
 it can be shown that the exponent in the degree
distribution becomes independent of $\varphi \left(\omega \right) $ 
if the
tail of $\varphi $ decays faster then any power law. 

Let $V_{N}=\left\{ 1,...,N\right\} $ be the vertex set of a random graph
space. We are mainly interested in the asymptotic properties for $N$ being
very large. We assign i.i.d. to each element $x$ from the set $V_{N}$ a
continuous positive real random variable (r.v.) $\omega \left( x\right) $
taken from a distribution with density function $\varphi \left( \omega
\right) $. The variable $\omega $ can be interpreted as a parametrization
of $V_{N}$. For a set 
\begin{equation}
\label{set_C}
C_{\,\omega _{0},\omega _{1}}=\left\{ x:\omega \left(
x\right) \in \left[ \omega _{0},\omega _{1}\right] \right\},
\end{equation}
we obtain 
\begin{equation}
\label{exp_set_C}
{\mathbb E}\left( \sharp C_{\,\omega _{0},\omega _{1}}\right)\,\, =\,\,N\cdot
\int\limits_{\omega _{0}}^{\omega _{1}}\varphi \left( \omega \right) d\omega 
\end{equation}
where $\sharp C_{\,\omega _{0},\omega _{1}}$ denotes the cardinality 
of the set (\ref{set_C}).
Without loss of generality,
 we always assume that $\varphi >0$ on $\left[
0,\infty \right),$ and that the 
 the tail of the distribution for $\varphi $
is a monotone function, namely
that 
$\varphi \in C^{2}\left( \left[ 0,\infty \right) \right)$ 
 and the second derivatives,
$D^{2}\left( \varphi ^{\mu }\right), $
 have no zeros for
$\left| \mu \right| \in \left( 0,\mu _{0}\right) $
 and 
$\omega >\omega _{0}\left( \mu \right).$

Edges are created by a directed process in which the basic events are
choices made by the vertices. All choices are assumed to be i.i.d. The
number of times a vertex $x$ makes a choice is itself a random variable
which may depend on $x$. We call this r.v. $d_{out}\left( x\right) $. The
number of times a vertex $x$ was chosen in the edge formation process is
called the in-degree $d_{in}\left( x\right) .$ Each choice generates a
directed edge. We are mostly interested in the corresponding undirected
graph. If we speak in the following about out-degree and in-degree we refer
just to the original direction in the edge formation process. Let 
$$
p_{\omega
}\,\,=\,\,\Pr \left\{ x\rightarrow y\mid \omega =\omega \left( y\right) \right\} $$
be the basic probability that a vertex $y$ with a fixed value of $\omega $
is chosen by $x$ if $x$ is about to make a choice. For a given realization $%
\xi $ of the r.v. $\omega $ over $V_{N}$ we assume: 
\begin{equation}
p_{\omega }\left( \xi ,N\right)\,\, =\,\,\frac{1}{\,\,N\,\,}\cdot \frac{A\left( \xi ,N\right) }{\,\,\left[
\varphi \left( \omega \right) \right] ^{\alpha }\,\,}.
\end{equation}
 where $\alpha \in \left( 0,1\right) $ and $A\left( \xi ,N\right) $ is a
normalization constant. It is easy to see that the condition 
$$
\int\limits_{0}^{\infty }\left[ \varphi \left( \omega \right) \right]
^{1-\alpha }d\omega\,\, <\,\,\infty 
$$ 
is necessary and sufficient to get 
$$
A\left(
\xi ,N\right) \rightarrow A\,\,>\,\,0,
$$ 
for $N\rightarrow \infty $ where convergence
is in the sense of probability. 
Therefore, we need $\alpha <1.$

One might
argue that the choice probabilities should depend more explicitly on the
actual realization $\xi $ of the r.v. $\omega $ over $V_{N}$ -not only via
the normalization constant. The reason not to do so is twofold. First it is
mathematically unpleasant to work with the empirical distribution of $\omega 
$ induced by the realization $\xi $ since one had to use a somehow
artificial $N-$ dependent coarse graining. Second the empirical distribution
is not really ''observed''\ by the vertices (having in mind for instance
individuals in a social network). What seems to be relevant is more the
common believe about the distribution of $\omega$. In this sense our
setting is a natural one. 

The emergence of a power law distribution in 
the above settings is not a
surprise as it might seem for the first glance. 
The situation is best
explained by the following example. Let us take
$$
\varphi \left( \omega \right)\,\, =\,\,C\cdot
e^{-\omega }
$$ 
and define a new variable 
$$
\omega ^{\ast }\,\,=\,\,\frac{1}{\,\,\left[
\varphi \left( \omega \right) \right] ^{\alpha }\,\,}\,\,=\,\,\frac{e^{\omega \alpha }}{\,\,C^{\alpha }\,\,}.
$$
  The new variable $\omega ^{\ast }$ can be seen as the
effective parameter to which the vertex choice process applies.
 What is the
induced distribution of $\omega ^{\ast }$ ? 
With 
$$
F\left( z\right)\,\, =\,\,
\Pr
\left\{ \omega ^{\ast }<z\right\} 
$$ 
we obtain 
\begin{equation}
F\left( z\right)\,\, =\,\,\int\limits_{0}^{\frac{1}{\alpha }\ln C^{\alpha }\cdot
z}\varphi \left( \omega \right) d\omega\,\, =\,\,-\frac 1{\,\,z^{1/\alpha }\,\,}-C,
\end{equation}
and therefore the $\omega ^{\ast }-$ distribution 
$$
\phi \left( \omega ^{\ast
}\right)\,\, =\,\,\frac{1}{\alpha }\cdot \frac{1}{\left( \omega ^{\ast }\right) ^{1+{1}/{\alpha }}}.
$$ 
This is a power law distribution with an exponent
depending only on $\alpha.$

The detailed results on 
the degree-degree correlations,
the clustering coefficients, and the second moment of degree distributions
are discussed in \cite{Blanchard:2004}.

\section{The statistics of bursts in systems close to a threshold of instability}
\label{sec:FAT_TAILS_2}
\noindent

Systems driven by random processes at a threshold of stability
may exhibit a random switching of a signal between a quiescent
(stable) and a bursting (unstable) state. Such an
\textit{intermittent} behavior
 is observed over a broad class of
different systems in physics and nonlinear dynamics.
Depending on
the origin, the intermittent behavior either meets the
classification proposed by Pomeau-Manneville
 \cite{Pomeau:1980} (the I-III
types intermittency) or fits the features of the crisis-induced
intermittency \cite{Grebogi:1983}.
In both cases, the parameters of the
models are static.
Another example of intermittent behavior,
called {\it on-off intermittency} 
  has been introduced in \cite{Platt:1993}
and then observed numerically and experimentally,
\cite{Pikovsky:1984,Yu:1990,Venkataramani:1995,Venkataramani:1996,Hammer:1994, Zumofen:1993,Redondo:1996,Heagy:1994,Platt:1994,Qu:1996,Pierre:2000}.
 The mechanism for this intermittency type
relies on a random forcing of a bifurcation parameter through a
bifurcation point.

The ergodic properties of a system at the threshold of stability
can be partially characterized by the distribution of the
quiescent times (the durations of laminar phases) $P(t)$ where
$t\in\mathbb{N}$. Indeed, a complete characterization of the
statistical properties of the system will imply the knowledge of
residence times distribution for all the regions of the phase
space and not only of the laminar regions. However, the former is
the first important statistical indicator of such dynamics and
this is a reason why we focus at the quiescent times
distributions in the present study.

Depending on the particular type of intermittency exhibited by the
system, the statistics of this distribution can asymptotically
meet either exponential laws, or power laws of exponent $\gamma$.
Particularly, the power-law statistics for the quiescent times
distribution is claimed to be typical for the systems
demonstrating the on-off type intermittency, and the value of
exponent $\gamma$ depends in general from the nonlinearity
characteristic of the dynamical system considered \cite{Yang:1996}. For
example, in the experiments with  ion-acoustic instabilities in a
laboratory plasma \cite{Pierre:2000}, due to nonlinear effects, the
exponent of power law depends on the value of a control parameter.

In the present section, we discuss the net effect produced on the
statistics of laminar phases by the stochastic fluctuations of a
system state variable (a bifurcation parameter) near the fluctuating
threshold of stability (a bifurcation point). 

We do not
refer to any definite physical system displaying an intermittent
behavior. For the toy model which we introduce, the bifurcation
parameter and the bifurcation point are considered as random
independent variables. It is supposed that intermittency takes place
in the system when the process crosses  the
threshold value.

The control parameter of the system is  the number
 $\eta\in
[0,1]$,
which represents a relative frequency of fluctuation of the
threshold value: varying the parameter
$\eta$ amounts to modifying the relation between the characteristic time
scales of the threshold variable and those of the state variable.

At $\eta=0$ (when the state and the threshold variables have the same time scale)
the statistics of laminar phases is exponential, while at $\eta = 1 $ (the limiting case of quenched
threshold) it can be power law; for the intermediate values
 $0<\eta<1$, the statistics is mixed
becoming exponential for sufficiently large times.

  In general, the statistics of laminar phases depends on the statistics of the random system
state variable and threshold described by the probability distributions $F$ and $G$ respectively.
For many distributions $F$ and $G$,
the proposed model can be solved analytically.

\subsection{Systems at a threshold of instability}
\label{subsec:FAT_TAILS_2.1}

Let us suppose that the state of a system can be characterized by a real number $x \in [0, 1]$.
Another real number $y \in [0, 1]$ plays the role of a threshold of stability. The system is stable
as long as $x < y$ and exhibits a sudden transition to the irregular state otherwise ($x \geq y$).

We consider $x$ as a random variable distributed with respect to some given probability
distribution function 
$$
P\{x < u\} \,\,= \,\,F(u).
$$
 In an analogous way, the value of the threshold
$y$ is also a random variable distributed over the interval $[0, 1]$ with respect to some other
probability distribution function (pdf) 
$$
P\{y < u\}\,\, = \,\,G(u).
$$
 In general, $F$ and $G$ are
two arbitrary left-continuous increasing functions satisfying the normalization conditions
$$
F(0)\,\, =\,\, G(0)\,\, =\,\, 0, \quad F(1)\,\, =\,\, G(1)\,\, = \,\,1.
$$
Given a fixed real number $\eta \in[0,1]$, we define a discrete
time random process in the following way. At time $t=0,$ the
variable $x$ is chosen with respect to pdf $F$, and $y$ is chosen
with respect to pdf $G$. If $x\geq y$, the process continues
and goes to time $t=1$. At time $t\geq 1,$ the following events
happen:

\textbf{i}) with probability $\eta$, the random variable $x$ is chosen
with pdf $F$ but the threshold $y$ keeps the value it had at time
$t-1$. Otherwise,

\textbf{ii}) with probability $1-\eta,$ the random variable $x$ is chosen
with pdf $F,$ and  the threshold $y$ is chosen with pdf $G$.

If $x\geq y,$ the process ends; if $x<y,$ the process
continues and goes to time $t+1.$

Eventually, at some time step $t,$ when the state variable $x$
exceeds the threshold value $y,$ the process stops, and the
system destabilizes, so this integer value $t=T$ acquired in this
random process limits the duration of the episode of laminar
dynamics. In the laps of time, the system regains the composure
state, when $x<y,$ and the process starts again.

While studying the above model, we are interested in the
distribution of the duration of laminar phases $P_{\eta}(T;F,G)$
provided the probability distributions $F$ and $G$ are given and
the control parameter $\eta$ is fixed.

Even if in our model the state variable $x$ is treated as a random variable, what is really
important in what follows is the corresponding pdf $F$. It would in fact be possible to treat $x$
as a deterministic dynamical variable defined by the iterated images of a map of the interval
$[0, 1]$. In this case we would assume the existence of an invariant ergodic (Bernoulli) measure
$dF$, for which $x$ is a generic orbit.

It is also to be noticed that the model proposed resembles closely the
coherent-noise models \cite{Newman:1996,Sneppen:1997} discussed in concern
with a standard sandpile model \cite{Bak:1987}  in self-organized
criticality, where the statistics of avalanche sizes and
durations take power-law forms. No exact analytical results
concerning the coherent-noise models have been obtained so far.
The proposed toy model has not been discussed in the literature
before and, in principle, is much simpler than those discussed in
\cite{Newman:1996,Sneppen:1997} since it does not involve any spatial
dynamics  typical of such extended systems with quenched
memory as the original sandpile models.

\subsection{Distribution of residence times below the threshold}
\label{subsec:FAT_TAILS_2.2}

We are interested in the probability $P_{\eta}(T;F,G)$ that the
random process introduced in the previous section ends precisely
at time $T$ with a crossing of the   threshold, provided
the distributions $F$ and $G$ are given and $\eta$ is fixed. We
shall denote $P_{\eta}(T;F,G)$ simply as $P(T)$. A straightforward
computation shows directly from the definitions of Sec.~\ref{subsec:FAT_TAILS_2.1}
that
\begin{equation}
\label{ex00}
 P(0) \,\,=\,\,\int^1_0 dG(y)\left( 1- F(y)\right).
\end{equation}
For $T\geq 1$, the system can either stay below the threshold in
the laminar state ( a "survival") ({\it S}) or surmount the
threshold to a burst state (a "death")  ({\it D}).
Both events can take place either in the
correlated way (with probability $\eta$; see (i) in Sec.~\ref{subsec:FAT_TAILS_2.1}) (we denote them $S_c$ and $D_c$), or in
the uncorrelated way (with probability  $1-\eta$; see (ii) in section Sec.~\ref{subsec:FAT_TAILS_2.1}) ($S_u$ and $D_u$). For $T = 1$,
we have for example
\begin{equation}
\label{ex01}
\begin{array}{lcr}
 P(1) &= &  P[SD_c]+P[SD_u]\\
      &= & \eta\int_0^1 dG(y) {\ }F(y)\left(1-F(y)\right) \\
      & &  + \left(1-\eta\right)
  \int^1_0 dG(y) {\ }F(y)\int_0^1
dG(z)\left(1-F(z)\right) \\
      &= & \eta B(1)+(1-\eta)A(1)B(0).
\end{array}
\end{equation}
Similarly,
\begin{equation}
\label{ex02}
\begin{array}{lcl}
P(2) &=& \eta^2 B(2) +  \eta(1 -\eta)\left(A(1)B(1) + A(2)B(0) \right) \\
     & & +
(1 -\eta)^2A(1)^2B(0),
\end{array}
\end{equation}
where
where we have defined,
 for $n = 0, 1, 2, \ldots ,$
\begin{equation}
\label{A_n}
A(n)\,\, =\,\, \int^1_0\,
dG(y)F(y)^n
\end{equation}
and
\begin{equation}
\label{B_n}
\begin{array}{lcr}
B(n) &=&  \int^1_0\,
dG(y)F(y)^n\left(1-F(y)\right)\\
 &=& A(n)-A(n+1).
\end{array}
\end{equation}
It is useful to introduce the generating function of $P(T )$:
$$
 \hat{P}(s)\,\,=\,\,\sum_{T=0}^{\infty}s^TP(T).
$$
The generating property of the function $\hat{P}(s)$ is such that
\begin{equation}\label{inverse}
  P(T)\,\,=\,\,\left.\frac 1{\,\,T!\,\,}\frac{\,\,d^T \hat{P}(s)\,\,}{\,\,ds^T\,\,}\right|_{s=0}.
\end{equation}
Defining the following auxiliary functions
\begin{equation}
\label{auxiliary_00}
\begin{array}{lclccc}
p(l) & = &\eta^l A(l+1), &\mathrm{for} & l\geq 1,& p(0)=0, \\
q(l) & = &(1-\eta)^l A^{l-1}(1),&\mathrm{for} & l\geq 1,&q(0)=0,\\
r(l)& = &\eta^l\left[\eta B(l+1)+ (1-\eta) A(l+1) B(0)\right],&\mathrm{for} & l\geq 1,&r(0)=0,\\
\rho& = &\eta B(1)+ (1-\eta)A(1)B(0). & & &
\end{array}
\end{equation}
we find
\begin{equation}\label{PS}
 \hat{P}(s)\,\,=\,\, B(0)+\rho s+ \frac
 {s\left[
\hat{r}(s)+\rho\hat{p}(s)\hat{q}(s)+\rho A(1)\hat{q}(s)+A(1)\hat{q}(s)\hat{r}(s)
 \right]}{\,\,1-\hat{p}(s)\hat{q}(s)\,\,},
\end{equation}
where $\hat{p}(s),\hat{q}(s),\hat{r}(s)$ are the generating
functions of $p(l), q(l), r(l),$ respectively.

In the marginal cases of $\eta=0$ and $\eta=1$,
 the probability
$P(T)$ can be readily calculated.
 For $\eta=0,$ equations (\ref{auxiliary_00}) and (\ref{PS})
give
\begin{equation}\label{eta0}
\hat{P}_{\eta =0}(s)\,\,= \,\,\frac{B(0)}{\,\,1-s\,A(1)\,\,}.
\end{equation}
Applying the inverse formula (\ref{inverse}) to equation (\ref{eta0}),
we get
$$
  P_{\eta=0}(T)\,\,=\,\,A^T(1)B(0) \,\,=\,\,\left[
\int ^1_0 dG(y)F(y)
  \right]^T\int^1_0 dG(y)\left(1-F(y)\right).
$$
Therefore, in this case, for any choice of the pdf $F(u)$ and
$G(u),$ the probability $P(T)$ decays exponentially.

For $\eta=1$, equations (\ref{auxiliary_00}) and (\ref{PS}) yield
\begin{equation}\label{eta1}
  \hat{P}_{\eta=1}(s)\,\,=\,\,\hat{B}(s),
\end{equation}
so that
\begin{equation}\label{Peta1}
  P_{\eta=1}(T)\,\,=\,\,B(T)\,\,=\,\,\int_0^1 dG(y)F(y)^T\left(1-F(y)\right).
\end{equation}


\subsubsection{Some examples of decay in the correlated case $\eta=1$}
\label{subsubsec:03bis}

We have just seen that  the probability
$P(T)$ decays exponentially, in the uncorrelated case $\eta=0,$ for any choice of the pdf $F $ and $G $.

In the correlated case $\eta=1$, many different types of behavior
are possible, depending on the form of the pdf $F $ and $G $. We
will examine an important class of $F $ and $G $, for which
$P_{\eta=1}(T)$ can be explicitly computed from equation
(\ref{Peta1}). We will take $F $ and $G $ as absolutely continuous
with respect to the Lebesgue measure, with
\begin{equation}
\label{mod01}
\begin{array}{lclr}
dF(u) &=& (1+\alpha)u^{\alpha}\, du, & \alpha > -1, \\
dG(u) &=& (1+\beta)(1-u)^\beta\, du, &  \beta > -1.
\end{array}
\end{equation}
Here we recognize the family of invariant measures of a map of the
interval with a fixed neutral point \cite{Wang:1989}.

Equation (\ref{Peta1}) gives in this case:
$$
P_{\eta=1}(T) = \frac{\Gamma(2+\beta) \; \Gamma(1 + T (1+\alpha))}
    {\Gamma(2 + \beta + T (1+\alpha))}
- \frac{\Gamma(2+\beta) \; \Gamma(1 + (T+1) (1+\alpha))}
    {\Gamma(2 + \beta + (T+1) (1+\alpha))} \;.
$$

Using Stirling's approximation, we get for $T>>1$:
\begin{equation}
\label{Stirlings}
P_{\eta=1}(T)\,\, = \,\,\frac{(1+\beta)\; \Gamma(2+\beta)\; (1+\alpha)^{-1-\beta}}
    {T^{2+\beta}}\; \left( 1+0\left(\frac{1}{T}\right) \right)\;.
\end{equation}
For different values of $\beta$, the exponent of the threshold distribution,
we get all possible power law decays of $P_{\eta=1}(T)$. Notice that the
exponent $(-2-\beta)$ characterizing the decay of $P_{\eta=1}(T)$ is independent
of the distribution $F$ of the state variable.

We were not able to prove that the asymptotic decay of
$P_{\eta=1}(T )$ is algebraic for any choice of the distributions
$F$ and $G$; nevertheless, we have not found any counterexample
contradicting this conjecture. Let us consider in particular the
case of uniform $F$ (the results of this section suggest in fact
that what determines the decay of $P(T )$ is mostly the threshold
pdf $G$): $P_{\eta=1}(T )$ is then a particular case of a
Riemann-–Liouville integral, and we did not find any case of
non--algebraic decay for large $T$ in the tables
\cite{Erdelyi:1954}.

\subsubsection{Upper and lower bounds for $P(T )$ for any $\eta$}
\label{subsubsec:03bisbis}

We will use the fact that
\begin{equation}
\label{fact01}
A(1)^n \leq A(n)\leq A(1) \quad \mathrm{and}
\quad 0 \leq B(n) \leq A(1), \quad n= 1, 2,\ldots .
\end{equation}
The upper bound for $A(n)$ is trivial, since $0 \leq F(y) \leq 1$
for any $y \in [0, 1]$. The lower bound
is a consequence of Jensen's inequality, and of the fact that the function
$x \to x^n$ is convex on
the interval $]0, 1[$ for any integer $n$.

We now replace these bounds for $A(n)$ and $B(n)$ in the general formula for $P(T)$ and the resulting
expressions in all the terms of the sums, except the one corresponding
 to the index $n = 0$ in
$P_I (T )$.
(This term, which corresponds to a sequence of correlated survivals, has to be treated
separately, in order not to lose information on the case $\eta = 1.$) Labelling by the index $k$ the
number of uncorrelated survivals in the sequence of events considered in the sum for $P(T)$, we get
\begin{equation}
\label{estim0a}
\begin{array}{lcl}
P_\eta(T )& \leq& \left[\eta^T B(T ) + \eta^{T-1}(1 -\eta )A(T )B(0)\right] \\
& +& \left[\eta A(1) + (1 -\eta)A(1)B(0)\right]
\sum_{k=1}^{T-1}\gamma_k^{T-1}\left[(1-\eta)A(1)\right]^k\eta^{T-1-k}
\end{array}
\end{equation}
and
\begin{equation}
\label{estim0b}
\begin{array}{lcl}
P_\eta(T )& \geq&
\left[\eta^T B(T ) + \eta^{T-1}(1 - \eta)A(T )B(0)\right] \\
& +& (1 -\eta)A(1)B(0)\sum_{k=1}^{T-1}\gamma_k^{T-1}
\left[(1-\eta)A(1)\right]^k\eta^{T-1-k}
\end{array}
\end{equation}
where $\gamma^{T-1}_k$
represents the number of sequences of $T -1$ events $c, u$ ($c$ = correlated survival,
$u$ = uncorrelated survival) containing a number $k$ of events $u$, so that
\[
\gamma_{k}^{T-1}\,\,=\,\,
\left(
\begin{array}{c}
T-1 \\ k
\end{array}
\right).
\]
This implies the upper bound
\begin{equation}
\label{estim0c}
\begin{array}{lcl}
P_\eta(T )& \leq&
\eta^T B(T ) + (1-\eta)
A(1)B(0)\left[
\eta+(1-\eta)A(1)
\right]^{T-1} \\
&+& \eta A(1)\left\{
\left[
\eta+(1-\eta)A(1)
\right]^{T-1}-\eta^{T-1}
\right\}
\end{array}
\end{equation}
and the lower bound
\begin{equation}
\label{estim0d}
\begin{array}{lcl}
P_\eta(T )& \geq&
\eta^T B(T ) + (1-\eta)A(1)^T B(0) \\
&=& \eta^T P_{\eta=1}(T ) + (1 - \eta)P_{\eta=0}(T ).
\end{array}
\end{equation}
We thus see that, for any $0 \leq \eta < 1$, the decay of distribution $P(T )$ is bounded by
exponentials. Furthermore, the bounds (\ref{estim0c}) and (\ref{estim0d}) are exact in the marginal cases $\eta = 0$ and
$\eta = 1$.

\subsubsection{Behavior of $P(T )$ for intermediate times}
\label{subsubsec:03bisbisbis}
\noindent

We have seen in Sec.~\ref{subsubsec:03bis}
that there exists a class of pdfs for which $P_\eta(T )$ decays like a
power law when $\eta = 1$. In section \ref{subsubsec:03bisbis},
 we show that for any $\eta < 1$ the asymptotic decay of
$P_\eta(T )$ is exponential.
 We now make some remarks about the behavior of $P_\eta(T )$ for
 $\eta$ close
to 1.
The first thing to be noted is that, for $T$ fixed, $P_\eta(T )$
is a continuous function of $\eta$, since it
is a finite sum of continuous functions (see Sec.~\ref{subsubsec:03bisbis}).
 This results, of
course,   imply that the continuity cannot be uniform in $T$. This
means that, for any fixed interval of times $[T_{-}, T_{+}]$, with
$T_{-}$ in the range of validity of the power-law asymptotes
(\ref{Stirlings}) of $P_1(T )$, $P_\eta(T )$ will be arbitrarily
close to the same power law for $\eta$ sufficiently close to $1$.
For times $T \gg T_{+}$, the decay becomes exponential. We shall
see in the next section that for the case of uniform densities, it
is possible to estimate the value of the crossover time to the
exponential behavior as a function of $\eta$.

\subsection{Distribution of quiescent times for the case of uniform densities}
\label{subsec:FAT_TAILS_2.3}
\noindent

In this section, we consider the distribution of quiescent times
for the special case of uniform densities $dF(u)=dG(u)=du,$ for all
$u\in [0,1]$ and for any $\eta\in [0,1]$.
In this case, simpler and implicit expressions can be given for
$\hat{P}(s)$ and $P(T)$.
  After some tedious but
trivial computation, we get from equation (\ref{PS}):
\begin{equation}\label{Psfg1}
  \hat{P}(s)\,\,=\,\,\frac 1{\,\,1+(1-\eta)\gamma(s)\,\,}\left[
  \frac{1+\gamma(s)}{s}-\eta\gamma(s)\right]\;,
\end{equation}
 where $\gamma(s)$ is defined by
\begin{equation}
\label{gamma_s}
\gamma(s)\,\,=\,\,\frac{\,\,\ln(1-\eta s)\,\,}{\eta s}\;.
\end{equation}
The asymptotic behavior of $P(T)$ is determined by the
singularity of the generating function $\hat{P}(s)$ that
is closest to the origin.

\begin{figure}[ht]
 \noindent
  \epsfig{file=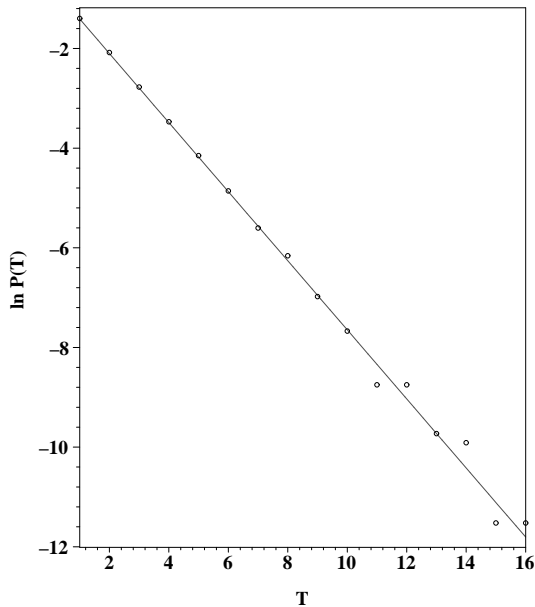, width=8cm, height =7cm, angle= -90}
 \label{Fig_FAT_TAILS_1}
\caption{Distribution of quiescent times for the uniformly distributed
variables $x$ and $y$. $P_{\eta}(T)$ decays exponentially
for $\eta=0$, consistently with the analytical result
$P(T)=2^{-(T+1)}$ (solid line).}
\end{figure}

For $\eta=0,$ the generating function has a simple pole
$\hat{P}(s)=(2-s)^{-1},$ and therefore $P(T)$ decays
exponentially, which agrees with the result of the previous
section. In Fig.~\ref{Fig_FAT_TAILS_1}, we have presented the distribution of
quiescent times $P(T)$ in log-linear scale for $\eta=0.$

For the intermediate values $1>\eta>0$, the generating function
$\hat{P}(s)$ has two singularities.
One pole, $s=s_0,$ corresponds
to the vanishing denominator $1+(1-\eta)\gamma(s),$ where $s_0$
is the unique nontrivial solution of the equation
\begin{equation}\label{s0}
  -\ln(1-\eta s)\,\,=\,\,s\,\frac{\eta}{\,\,1-\eta\,\,}\;.
\end{equation}
Another singularity, $s=s_1=\eta^{-1},$ corresponds to the vanishing argument of
the logarithm.
It is easy to see that $1< s_0< s_1,$
so that the dominant singularity of $\hat{P}(s)$ is of polar type, and
the corresponding decay of $P(T )$ is exponential, with rate $\ln\left(s_0(\eta)\right)$, for times much larger
than the crossover time $T_c(\eta) \sim \ln\left(s_0(\eta)\right)^{-1}.$

The results of Sec.~\ref{subsubsec:03bisbis} about the upper bound for the distribution $P(T )$ allow us to be
more precise about this decay rate.
In particular, since $B(T ) \leq A(1)$, it follows
from (\ref{estim0c})
that
\begin{equation}
\label{bound001}
P(T )\,\, \leq \,\, \frac{\,\,\eta A(1) + (1 - \eta)A(1)B(0)\,\,}{\,\,\left(\eta + (1 - \eta)A(1)\right)^{T-1}\,\,}
\end{equation}
which in the case of uniform densities gives
\begin{equation}
\label{bound002}
P(T )\,\,\leq \,\, \frac 12\,\left( \frac{\,\,1+\eta\,\,}{2}\right)^T.
\end{equation}
We have then
\begin{equation}
\label{bound003}
\frac 1{\,\,s_1\,\,}\,=\,\eta\,<\,\frac 1{\,\,s_0\,\,}\,\leq\,\frac{\,\,1+\eta\,\,}2
\end{equation}
and we see that the rate $\ln\left(s_0(\eta)\right)$
vanishes like $1-\eta$ as $\eta$ tends to $1$.

\begin{figure}[ht]
 \noindent
 \epsfig{file=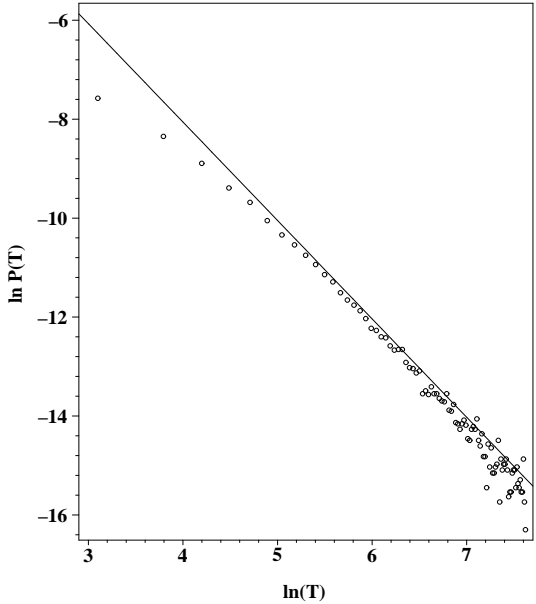, width=8cm, height =7cm,  angle= -90}
\label{Fig_FAT_TAILS_2}
\caption{ Distribution of quiescent times for the uniformly distributed
variables $x$ and $y$. We show the power-law decay of $P_{\eta=1}(T)$
plotted in the log-log scale. The solid line is given by  (\ref{PTeta1}).}
\end{figure}

When $\eta$ tends to $1$, the two singularities $s_0$ and $s_1$ merge. More precisely, we have
\begin{equation}\label{Pseta1}
\hat{P}_{\eta=1}(s)=\frac{s+(1-s)\ln(1-s)}{s^2}.
\end{equation}
The corresponding dominant term in (\ref{Pseta1}) is of order
$\mathrm{O}(T^{-2})$ \cite{Flajolet:2008}.
This obviously agrees with the exact result we get from
equation
(\ref{Peta1}), with $dF(u)=dG(u)=du:$
\begin{equation}\label{PTeta1}
P_{\eta=1}(T)\,\,=\,\,\frac {1}{\,\,(T+1)(T+2)\,\,}\;.
\end{equation}
In Fig.~\ref{Fig_FAT_TAILS_2}, we have drawn the distribution of quiescent times
$P_{\eta=1}(T)$ that exhibits the power-law decay, with the slope
$\gamma=-2.$

\begin{figure}[ht]
  \epsfig{file=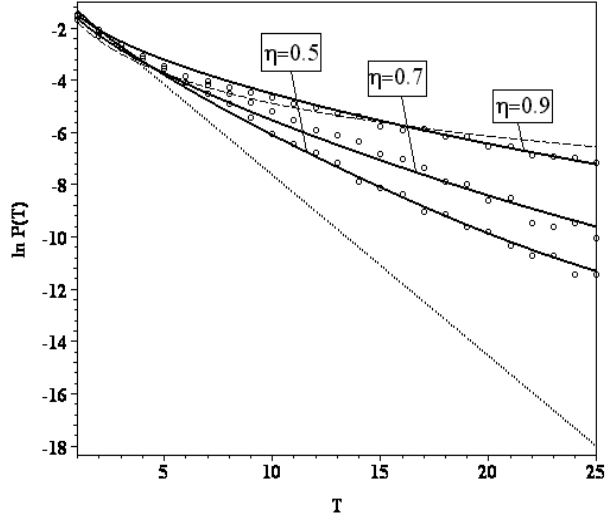, width=8cm, height =7cm}
\label{Fig_FAT_TAILS_3}
\caption{Distribution of quiescent times for the uniformly distributed
variables $x,$ $y$ at the intermediate values $\eta=0.5$,
$\eta=0.7$, $\eta=0.9$ (circles). For comparison, the dotted line
$2^{-T-1}$ presents the exponential decay  for $\eta=0$, and the
dashed line corresponds to $1/(T+1)(T+2)$  for $\eta=1$, see (\ref{PTeta1}).
The solid line is given by  (\ref{etavary1})
for $\eta=0.5$, $\eta=0.7$, $\eta=0.9$.}
\end{figure}

In the case of uniform densities, it is possible to get an
expression of $P(T)$ for all times, and for any value of
$\eta$, by applying the inversion formula (\ref{inverse}) to
(\ref{Psfg1}):
\begin{equation} \label{etavary1}
\begin{array}{ll}
P(T) = &\frac{\eta^T}{(T+1)(T+2)} \\
       & +
    \sum_{k=1}^{T} \frac{\eta^T}{(T-k+1)(T-k+2)\,k}\,\sum_{m=1}^{k}
    \left(\frac{1-\eta}{\eta}\right)^m  c(m,k)\;,
\end{array}
\end{equation}
where $c(m,k)$ is defined by
\begin{eqnarray*}
c(m,k) = m! \, \sum_{\begin{array}{c}
    \scriptstyle l_1 + l_2 + \cdots + l_{m} \,= \,k\\
        \scriptstyle l_i \ge 1
    \end{array}}  \hspace{10cm}\\
\frac{l_1 \, l_2 \, \cdots \, l_{m-1} \, l_m}
    {(l_1+1) \, (k-l_1) \, (l_2+1) \, (k-l_1-l_2) \, \cdots \,
    (l_{m-1}+1) \, (k-l_1-l_2-\cdots-l_{m-1}) \, (l_m+1)} \;.
\end{eqnarray*}
When $\eta\neq 0$, there is an alternative way of
writing the previous expression:
\begin{equation} \label{etavary2}
\begin{array}{ll}
P(T) = & \frac{\eta^T}{(T+1)(T+2)} \\
  & +
    \sum_{k=1}^{T} \frac{\eta^{T+1}}{(T-k+1)(T-k+2)}\,
    \sum_{l=1}^{\infty} (1-\eta)^l  \, b(l,k)\;,
\end{array}
\end{equation}
where $b(l,k)$ is defined by

$$
b(l,k) = \sum_{\begin{array}{c}
    \scriptstyle i_1 + i_2 + \cdots + i_{l} \,= \,k\\
        \scriptstyle i_j \ge 0
    \end{array}}
\frac{1}{(i_1+1) \, (i_2+1) \, \cdots \, (i_{l-1}+1) \, (i_l+1)} \;.
$$
In Fig.~\ref{Fig_FAT_TAILS_3},
we have plotted the distribution of quiescent times
$P_{\eta}(T)$ for the intermediate values $\eta = 0.5$, $\eta =
0.7$, $\eta = 0.9$, together with the analytical result
(\ref{etavary1}).

Note that in Fig.~\ref{Fig_FAT_TAILS_1} and Fig.~\ref{Fig_FAT_TAILS_3} (where $\eta \neq 1$), we only plot
distributions $P(T)$ up to relatively short quiescent times
($T=16$, $T=25$), since these times are already greater than the
crossover time $T_c(\eta)\sim 1/\ln(s_0)$ to the exponential
decay $\exp{(-\,\ln(s_0)\,T)}$ ($s_0$ defined by equation
(\ref{s0})). For much longer times, very few survivals are
observed, and the statistics gets bad. Of course, $T_c(\eta)$
grows as the parameter $\eta$ tends to 1, so that we have good
statistics for longer and longer times (in Fig.~\ref{Fig_FAT_TAILS_2},
for $\eta =1$, the plot is for quiescent times $20 \le T \le 2000$).

\bigskip

A natural question arising in this context is about the relationship
between the ergodic invariants that quantify the dynamics of
deterministic systems, for example the Lyapunov exponents, and the
scaling laws. The corresponding question for
models of self--organized criticality is certainly also pertinent
since in that case a relation is known between the Lyapunov
spectrum and the transport properties \cite{Blanchard:2000}. In
our case, however, because of the dynamical character not only of
the state variable but also of the threshold, some extension of
the definition of the invariants would be needed, which is beyond
the scope of our discussion.

\section{Fat tails in queuing systems}
\label{sec:FAT_TAILS_3}
\noindent

In a simplified model of human activity, 
\cite{Barabasi:2005,Vazquez:2005,Vazquez:2006},
it is viewed  as a decision based {\it queuing system}
 (QS)
where tasks to be executed arrive randomly and accumulate before
a server ${\cal S}.$
Under
the priority-based scheduling rules,
in which 
each incoming task is endowed with a {\it priority
index}  (PI) indicating the  urgency to process the job,
the timing
of the tasks follows fat tails probability distribution, (i.e the
activity of the server exhibits bursts separated by long idle
periods with the
ubiquitous Poisson behavior).

There are two types of
dynamics:
\begin{itemize}
    \item[{\bf i).}] \vspace{0.1cm} {\it Service policies based on fixed priority indexes}. This case which
    is considered in \cite{Barabasi:2005,Vazquez:2005,Vazquez:2006}
     assumes that the value $a$ of the PI  is fixed once for all. Accordingly, very low
    priority jobs are likely to never be served. To circumvent this
    difficulty \cite{Barabasi:2005,Vazquez:2005,Vazquez:2006}  introduce an ad-hoc probability
    factor $0\leq p\leq 1$ in terms of which the limit
    $p \rightarrow  1$ corresponds  to a deterministic scheduling strictly based
    on the
   PI's while in the other limit $p \rightarrow  0$ the purely random scheduling is in use.

 In this setting,
    the {\it waiting time distribution}
 (WTD) of the tasks before
    service is shown to asymptotically exhibit a fat tail
    behavior. The main point of the  Barabasi's contribution  is to show that
    { PI-based scheduling rules can alone generate fat tails in the WTD of unprocessed jobs}.

   \item[{\bf ii).}]  \vspace{0.2cm} {\it Service policies based on  time-dependent priority indexes}. Here the priority
    index is {time-dependent}. This typically models situations where {\it the urgency to
      process a task increases with time} and $a(t)$ will hence be represented by increasing time functions.

     Clearly, scheduling rules based on such a time-dependent PI do offer
     new specific  dynamical features. They are
      directly relevant in several contexts such as:

     \begin{itemize}
     \item[]  a). {\it Flexible manufacturing systems with limited resource}. Here a single server is
    conceived  to process different types of jobs but only  a single type can be
produced at a given time $t$
        (i.e. this is the limited resource constraint). Accordingly, the basic
problem is to dynamically schedule the production to optimally
match random demand arrivals for each types of items. The dynamic
scheduling can be optimally achieved by using time-dependent
priority indexes ({\it the Gittins' indexes}) which 
specify in
real time, which type of production to engage
\cite{Dusonchet:2003}. Problems of this type belong to the wider class referred as
the {\it Multi-Armed Bandit Problems}
in operations research.

\item[]  b). {\it Tasks with deadlines}.  This situation,
 can be idealized by a queuing system where each incoming item has a deadline before
 which it definitely must be processed, \cite{Lehozcky:1996,Lehozcky:1997,Doytchinov:2001}. In this case,
 to be later  discussed in the present paper, we can explicitly derive
 the lead-time profile of the waiting jobs obtained under  several scheduling
rules, including the (optimal) time-dependent priority rule  known
as the {\it earliest-deadline-first}
 policy.

\item[]  c). {\it Waiting time-dependent feedback queuing
systems}. In queuing networks, priority indexes based on the
waiting times can be used to schedule the routing through the
network. For networks with loops, such scheduling policies are
able to generate generically stable oscillations of the
populations contained in the waiting room of the queues, 
\cite{Filliger:2005}.
     \end{itemize}
\end{itemize}

 In the context of QS, the {\it waiting time probability
distribution} (WTD), (i.e. the time the tasks spend in the queue
before being processed) is a central quantity to  characterize the
dynamics. It
  strongly depends on the arrival and service stochastic
processes - in particular to the distributions of the {\it
inter-arrival} and {\it service} time intervals. The first moments
of these distributions, enable to define the  traffic load 
$$
\rho
\,\,=\,\, \frac{\lambda}{\mu} \,\,\geq \,\,0,
$$ 
(i.e.  the ratio between the mean
service time ${1 / \mu}$ and the mean arrival time ${1 /
\lambda}$) and clearly the stability of elementary QS is ensured
when $0 \leq \rho <1$. Focusing on the WTD,
\cite{Barabasi:2005,Vazquez:2005,Vazquez:2006} emphasized that
heavy tails in the WTD can  have several origins, three of which
are listed below:

\begin{itemize}
    \item[]  1). the {\it heavy traffic load of the server} which induces large
  "bursty" fluctuations in both the WTD and the busy period (BP) of the
  QS. For QS with feedback control driving the dynamics to heavy
  traffic loads, this allows to generate self-organized critical (SOC) dynamics, \cite{Blanchard:2004}
  and the resulting fat tails distribution exhibit a decay following a
  $-3/2$ exponent.

    \item[] 2). the presence of {\it fat tails in the service time distribution} produce fat tails of the
    WTD a property which is here independent of the scheduling
    rule \cite{Boxma:2004}. For the convenience of the reader, we give here a short review of these  results.

    \item[] 3). priority index scheduling rules as discussed
    in \cite{Barabasi:2005,Vazquez:2005,Vazquez:2006}.
\end{itemize}

  In this section we pay the essential attention to the case iii) but contrary to the discussion carried in
  \cite{Barabasi:2005,Vazquez:2005,Vazquez:2006},
  we shall here consider the dynamics in presence of {\it age-dependent priority
  indexes}. 
As it could have been expected, these aging mechanisms
  generate new behaviors that will be explicitly discussed for two
  classes of models.

\subsection{Waiting time distributions for queuing systems with fat tail service times}
\label{subsec:FAT_TAILS_3_1}

Let us reproduce here a result recently
derived in \cite{Boxma:2004}.

\begin{theorem}[Boxma]
\label{theorem_Boxma}
Assume that the
(random) service time in a $M/G/1$ QS is drawn from a PDF with a
regularly varying tail at infinity with index $\nu \in (-1, -2)$,
(regularly varying  with index $\nu \in (-1, -2)$ $\Rightarrow$
fat tail with index $\nu \in (-1, -2)$). For this range of
asymptotic behaviors of the PDF, the first moment $\beta$ of the
service exists.

Assume further that the service is delivered
according to a random order discipline. Then the waiting time
distribution $W_{{\rm ROS}}$ exhibits a fat tail with index
$\left(1-\nu \right)\in (-1,0)$ and more precisely, we can write
\begin{equation}\label{ROS}
\Pr(W_{{\rm ROS}} > x )\,\,\propto\,\, C\,\frac{\rho}{\,\,1-\rho\,\,}\,
    \frac{\,\,h(\rho, \nu)\,\,}{\beta \, \Gamma(2-\nu)}\,  x^{1-\nu}\, {\cal
    L}(x),
\end{equation}
where $\rho <1$ is the traffic intensity, $\beta$ the
average service time, ${\cal L}(x)$ a slowly varying function and
$$
h(\rho, \nu) = \int_{0}^{1} f(u, \rho, \nu) du,
$$
with:
$$
\begin{array}{ll}
f(u, \rho, \nu) =& \frac{\rho}{\,\,1-\rho\,\,}\left(\frac{\rho \, u}
{\,\,1-\rho\,\,}\right)^{\nu\,-1} \,\left(1-u\right)^{{1 / (1-\rho)}}\\
 &  +
\left(1+ \frac{\rho}{\,\,1-\rho\,\,} \right)^{\nu}\,
\left(1-u\right)^{{1/ (1-\rho)}-1}.
\end{array}
$$
\end{theorem}

The fat tail behavior given in
 (\ref{ROS}) is therefore entirely inherited from the  fat tail
behavior of the service and is not affected by any reduction of
the traffic intensity $\rho$. Note also that change of the
scheduling rule cannot get rid of this fat tail behavior. This
point can be explicitly observed in \cite{Cohen:1973,Pakes:1975},
who show  that for the previous $M/G/1$ QS with a random order
service (ROS) service discipline, one obtains:
\begin{equation}\label{FCFS_ROS}
\Pr(W_{{\rm ROS}} > x )\,\,\propto_{x \to \infty}\,\, h(\rho,\nu) \cdot \Pr(W_{{\rm FCFS}} > x ), 
\end{equation}
\noindent from which we directly observe that {\it the fat tail in
the asymptotic behavior in not altered by a change of the
scheduling rule}.

Note finally that for the $M/M/1$ QS, (i.e.
exponential service distributions and hence no fat tail), \cite{Flatto:1977}
 shows that the random order service scheduling rule yields:
\begin{equation}\label{ROS-MM1}
\Pr(W_{{\rm ROS}}>x)\,\, \propto_{x \to \infty}\,\, C_{\rho} x^{-{5 / 6}}\,
e^{- \gamma x- \delta x^{{1/ 3}}},
\end{equation}
with
$$
C(\rho) = 2^{2/3} 3^{-{1/2}} \pi^{{5/6}} \rho^{{17
/ 12}} \frac{\,\,1+ \rho^{1/ 2}\,\, }{\left[1-\rho^{1 /
2}\right]^3} \,\,\exp\,\left\{\frac{1+ \rho^{{1/2}}}{1-\rho^{{1/2}}}\right\},
$$
and
$$
\gamma\,\,=\,\, \left(\rho^{-{1
/2}}-1\right)^{2}\quad {\rm and }
\quad \delta =3\left[\frac {\,\pi\,}2\right]^{{2/
3}} \rho^{-{1 /6}},
$$
 which has to be compared with the FCFS scheduling
discipline, which for the same $M/M/1$ QS reads as, \cite{Cohen:1973}:
\begin{equation}\label{ROS-MM2}
\Pr(W_{{\rm FCFS}}> x)\,\,=\,\, \frac 1\beta\, (1-\rho) e^{-
 (1-\rho)x/\beta}.
\end{equation}
 While the  detailed behaviors given by
 (\ref{ROS-MM1}) and (\ref{ROS-MM2})clearly differ, they however
both share, in accord with \cite{Barabasi:2005}, an exponential decay.

\subsection{Scheduling based on time dependent priority indexes}
\label{subsec:FAT_TAILS_3_2}

  The most naive approach to discuss the dynamics of QS
with scheduling based on {\it time-dependent priority indexes} is
to think of a population model in which the members suffer aging
mechanisms which ultimately  will kill them. 

Naively, we may
consider the population of a city in which members are either born
in the city or  immigrate into it at a certain age and finally die
in the city. Assuming that the death probability {\it depends on
each individual age}, the study of the age structure of the
population exhibits some of the salient features of our original
QS. 

First, we discuss this class of models
and then
return to the
original model of L. Barab{\'a}si \cite{Barabasi:2005}
to consider a simple QS where each
task waiting to be processed carries a deadline (playing the role
of a PI) and as time flows the these deadlines steadily reduce -
this implies a ({\it time dependence of the PI}). At a given time,
the scheduling of the tasks follows the "earliest-deadline-first"
(EDF) policy and given a queue length configuration, we shall
discuss the lead-time (lead-time = deadline - current time)
profile of the tasks waiting to be served.

\subsubsection{Tasks population dynamics with time dependent priority indexes}
\label{subsubsec:POPU}

 Consider a population of tasks waiting to be processed
by ${\cal S}$ with the following characteristics:

\begin{itemize}

\item[] i). An inflow of  new tasks steadily enters into the
queuing system. Each tasks is endowed with a priority index (PI)
$a$ which indicates its degree of urgency to be processed. In
general, the tasks are heterogenous as the PI are different. In
the time interval $[t, t+\Delta t]$, the number of incoming jobs
exhibiting an initial PI in the interval $[a, a + \Delta a]$ is
characterized by $G(a,t) \Delta t \Delta a$.

\vspace{0.3cm}

\item[] ii). Contrary to the
    situations  discussed in \cite{Barabasi:2005}, an "aging" process
    directly affects the urgency to process a given task. In other
    words the priority index $a$ is not frozen in time but $a=a(t)$
    monotonously increases with time $t$. For an infinitesimal time increase $\Delta t$, in the simplest
    case  we shall have
    $$
a(t+ \Delta t)\,\, =\,\, a(t) +  \Delta t.
$$ 
Here we slightly  generalize this and allow inhomogeneous aging
    rates written as $p(a)>0$ meaning that 
$$
a( t+ \Delta t)\,\, =\,\, a(t)
    + p(a) \Delta t.
$$

\vspace{0.2cm}

\item[] iii). The scheduling policy  depends on the PI of the tasks
in the queue and we will focus on the natural policy {\it "process
the highest PI first"}.

\vspace{0.3cm}

    \item[] iv). at time $t$, a scalar field $M(a,t)$ counts
 the number of waiting tasks with priority
index $a$. Hence $M(a,t) \Delta a$ is the number with PI $p(a) \in
[a + \Delta a]$. Accordingly, the total workload facing the human
server $\cal {S}$ will be given, at time $t$ by:

\begin{equation}\label{WL}
    L(t)\,\, = \,\,\int_{0}^{\infty} M(a,t) \,da.
\end{equation}

\item[]   v).  In the time interval  $[t, t+\Delta t]$, the server
$\cal{S}$  processes tasks with an $a$-dependent rate $\mu(a)
\Delta t$. Typically $\mu(a)$ could be a monotonously increasing
function of $a$. As the service rate $\mu(a)$ explicitly depend on
the PI $a$, it therefore plays an effective role of service
discipline.

\end{itemize}

\noindent
 The previous elementary hypotheses imply an evolution in the form:
  $$
\begin{array}{lll}
M(a + p(a)\Delta t, t + \Delta t) \Delta a &\approx &  M(a,t)\, \Delta a- \mu(a)\, M(a,t) \Delta a \Delta t \\
 & &+ G(a,t)\,\Delta t \Delta a.
\end{array}
$$

    \noindent
Dividing by $\Delta a  \Delta t$, we end, in the limits $\Delta a
\rightarrow 0$ and $\Delta t \rightarrow 0$,  with the scalar
linear field  equation:

\begin{equation}\label{EVOL}
  \frac{\partial}{\partial
    t}M(a,t) + p(a)\frac{\partial}{\partial
    a}M(a,t)+ \mu(a) M(a,t)\,\, =\,\, G(a,t).
\end{equation}

\noindent It is worth to remark that the dynamics given by
(\ref{EVOL}) is closely related to the  famous McKendrick's age
structured population dynamics, \cite{Brauer:2001}.

\noindent Assuming stationarity for the incoming flow of tasks
(i.e. $G(a,t)= G_{s}(a)$), the linearity of (\ref{EVOL})
enables to explicitly write its stationary solution as:

\begin{equation}\label{STA}
    M(a) = \pi(a) \left[C + \int_{0}^{a}  \frac{G_{s}(z)}{\,\,p(z) \,
    \pi(z)\,\,}\, dz\right],
\end{equation}

\noindent where
\begin{equation}\label{pi}
    \pi(z) = \exp\left\{-\int^{z}\frac{\mu(y)}{
    \,\,p(y)\,\,}\, dy\right\},
\end{equation}

\noindent with an integration constant $C < \infty$ remaining yet
to be determined. Assume that the  PI attached to the incoming
jobs do not exceed a limiting value $T$, namely:
\begin{equation}\label{T}
    G(a,t) = \mathbb{I}(a < T)\, \hat{G}(a,t)\quad \Rightarrow \quad
    G_{s}(a) = \mathbb{I}(a < T) \,\hat{G}_{s}(a),
\end{equation}

\noindent where $\mathbb{I}(a < T)$ is the indicator function. In
other words  (\ref{T})  indicates that the new coming jobs do
not exhibit arbitrarily high PI's.

 \noindent This enables to define:

\begin{equation}\label{NOTA}
\Psi(T)\,\, = \,\,\int_{0}^{\infty} \frac{G_{s}(z)}{\,\,p(z) \,
    \pi(z)\,\,}\, dz \,\,= \,\,\int_{0}^{T} \frac{\hat{G}_{s}(z)}{\,\,p(z) \,
    \pi(z)\,\,}\, dz
\end{equation}
\noindent and  (\ref{STA}) reads as:
\begin{equation}\label{FINASOL}
M(a) = \left\{\begin{array}{c}  \pi(a) \left[C + \int_{0}^{a}
 \frac{\hat{G}_{s}(z)}{\,\,p(z) \,
    \pi(z)\,\,}\, dz\right]\qquad\mbox{if} \qquad a\leq T
    \\ \pi(a) \left[C+ \psi(T)\right]\qquad \qquad \qquad\mbox{if} \qquad a>
    T.
\end{array}\right.
\end{equation}
\noindent The asymptotic behavior  of $M(a)$ for  $a \rightarrow
\infty$ is entirely due to $\pi(a)$, (the square bracket terms are
bounded by constants) and therefore   (\ref{pi}) and
(\ref{FINASOL}) imply:
\begin{equation}\label{ASY}
    M(a) \approx \pi(a) \approx
    \left\{\begin{array}{c}
e^{{-k / q} a^{q}}\qquad\mbox{when} \qquad \frac{\mu(a)}{p(a)}
\propto ka^{q-1},\qquad q >  0,\\
{a^{-k}}\qquad\,\,\,\,\,\,\,\,\,\mbox{when} \qquad
\frac{\mu(a)}{p(a)} \propto  k/a, \qquad \qquad
\qquad\qquad\qquad \\
\end{array}
    \right.
\end{equation}
\noindent In view of  (\ref{ASY}), the following alternatives
occur:
\begin{itemize}
    \item[] a). For  $q < 0,$ in  (\ref{ASY}), the
integral $\int_{0}^{\infty}M(z)\, dz$ does not exist. In this case
an ever growing population of tasks accumulates in front of the
server and the queuing process is exploding.

\vspace{0.3cm}

    \item[] b). For $q > 0$, a stationary regime exists and in this case the constant
$C < \infty$ in  (\ref{FINASOL}) can be determined by solving:

\begin{equation}\label{BALANCE}
    \int_{0}^{\infty} G_{s}(z) \, dz = \int_{0}^{\infty} M(z)
    \mu(z) \,dz,
\end{equation}

\noindent which  expresses a global balance between the stationary
incoming and out going flows of tasks.

\vspace{0.3cm}

    \item[] c). 
For $q=0$ which implies that 
$$
\frac{\mu(a)}{p(a)}\,\, \propto\,\, \frac{  k}{a},
$$
 (\ref{ASY}) {\it produces an exponent-$k$ fat tail distribution
for $M(a)$ counting the number of waiting tasks with PI $a$} in
the system. For $T < \infty$ and $ a \rightarrow \infty$, the fat
tail of $M(a)$  is populated by long waiting  tasks i.e. those
having spent more than $a-T$ waiting inside the system before
being served. In the limiting case, for which 
$$
\mu(a) \,\,=\,\, \mu \,\,=\,\, {\rm
const}
$$ 
and $p(a) =a$ (i.e. aging directly proportional to time)
which leads to $q=0$ in  (\ref{ASY}), the density $M(a)$
coincides directly with the WTD for $a \rightarrow \infty$.
\end{itemize}

\noindent This population model shares  several features  with the
Barab{\'a}si's model \cite{Barabasi:2005}, namely:

\begin{itemize}
    \item[] a). When a stationary regime exists, the function
$\hat{G}_{s}(a)$ which  here plays the role of the initial PI
distribution in \cite{Barabasi:2005,Vazquez:2005,Vazquez:2006}, does not affect the tail behavior given by
 (\ref{ASY}).

\vspace{0.3cm}

    \item[] b).  The scheduling rule here is implicitly governed  by the service rate
    $\mu(a)$ which itself depend on time as the PI $a=a(t)$ are time-dependent.
    Note that $\mu(a)$ directly influences the asymptotic behavior of (\ref{ASY}). In particular for case
    c), the tail exponent explicitly depends on  $\mu(a)$.

\end{itemize}

\noindent Besides  the similarities, we now also point out the
important differences between the present population model and the
model discussed in  \cite{Barabasi:2005,Vazquez:2005,Vazquez:2006}:

\begin{itemize}

    \item[] a) the service is not restricted to a single task at
a given time (i.e. the service resource is not limited). Indeed
$\mu (a)$ describes an average flow of service and hence several
tasks can be processed simultaneously - (in the city population
model the service corresponds to death and several individual may
die simultaneously).

\vspace{0.3cm}

    \item[] b) while the fat tail in \cite{Barabasi:2005,Vazquez:2005,Vazquez:2006} is entirely due to the scheduling rule and
therefore occurs even for QS far from traffic saturation, this is
not so in the population model. Indeed in this last case, fat
tails are due to heavy traffic loads occurring when the flow of
incoming tasks nearly saturates the server, (this is implied by
$q=0$ in(\ref{ASY})) - for lower
 loads occurring when $q>0$  the fat tail in (\ref{ASY})
 disappears.

\end{itemize}

\subsubsection{Stochastic dynamics. Real-time queuing dynamics}
\label{subsubsec:TIMERQ}

 In this section we will use  the results of the
real-time queuing theory (RTQS), pioneered in \cite{Lehozcky:1996}, to
explore situations where the incoming jobs have a deadline - this
problem is already suggested in \cite{Barabasi:2005}. Based on
\cite{Lehozcky:1996,Lehozcky:1997,Baldwin:2000} and
\cite{Doytchinov:2001},  first recall the basic hypotheses and
the relevant results of RTQS's. Consider a general single server
QS with arrival and service being described by independent renewal
processes with average ${1 / \lambda}$ respectively ${1 /
\mu}$ and finite variances for both renewal processes. Each
incoming task arrives with a random hard time relative deadline
${\cal D}$ drawn from a PDF $G(x)$ with a density $g(x)$:
$$
\Pr\left\{ 0\leq {\cal D} \leq x \right\} = G(x),
$$
 with average $\langle {\cal D } \rangle$:
$$
\langle{\cal D}\rangle = \int_{0}^{\infty}\left(1 -
    G(x)\right)dx = \int_{0}^{\infty}x g(x)dx.
$$
 At a given time $t$, we define the lead time ${\cal L}$
to be given by:
\begin{equation}\label{LTIME}
    {\cal L}\,\, =\,\, {\cal D} - t,
\end{equation}
 Assume now that the lead time ${\cal L}$ plays the role
of a priority index and  the service is delivered by using the
{\it earliest-deadline-first}
 (EDF) rule with preemption (i.e.
the server always processes the job with the shortest lead time
${\cal L}$). Preemption implies that whenever an incoming job
exhibits  a shorter ${\cal L}$ than the one currently in service,
this incoming job is processed before,
 (i.e. {\it preempts}), the
 currently engaged task which service is postponed. The EDF rule
 directly
 corresponds  to the deterministic policy (i.e. $p=0, \gamma= 0$
 in the original Barab{\'a}si's contribution \cite{Barabasi:2005}.

 At a given
time, one can define a probability distribution corresponding to
the {\it lead time profile} (LTP),
$$
F(x)\,\,=\,\,\Pr\left\{-\infty\leq
{\cal L} \leq x\right\},
$$
of the jobs waiting in the QS. The
LTP specifies the repartition of tasks having a given ${\cal L}$
at time $t$. Knowing the queuing population $Q$ at a given time,
it is shown in \cite{Doytchinov:2001} that for heavy traffic
regimes, the LTP can, in a first order approximation scheme,
expressed by a simple analytical form.
Indeed, following
\cite{Doytchinov:2001}, let us define the
 frontier $\hat{{\cal F}}(Q)>0$ as
the unique solution of the equation

\begin{equation}\label{FRONTIER}
    \frac{Q}{\lambda}\,\, =\,\,\int_{\hat{{\cal F}}(Q)}^{\infty}\left(1 -
    G(x)\right)dx, \qquad \quad
x \in[l,\infty)\subset  \mathbb{R}^{+}.
\end{equation}
Let us also define the frontier $\mathcal{F}(Q)$ as
\begin{equation}
\label{new_frontier0}
\mathcal{F}(Q)\,\,=\,\,
\left\{
\begin{array}{lrr}
\hat{\mathcal{F}}(Q) & \mathrm{when} & Q\leq Q^{*},\\
\left( \langle \mathcal{D}\rangle -\frac{Q}{\,\,\lambda\,\,}\right)\leq l  & \mathrm{when} & Q > Q^{*},
\end{array}
\right.
\end{equation}
with $Q^*$ being defined by
$\hat{\mathcal{F}}(Q)=l$ where $l$ defined in (\ref{FRONTIER}).

\begin{figure}[ht]
     \epsfig{file=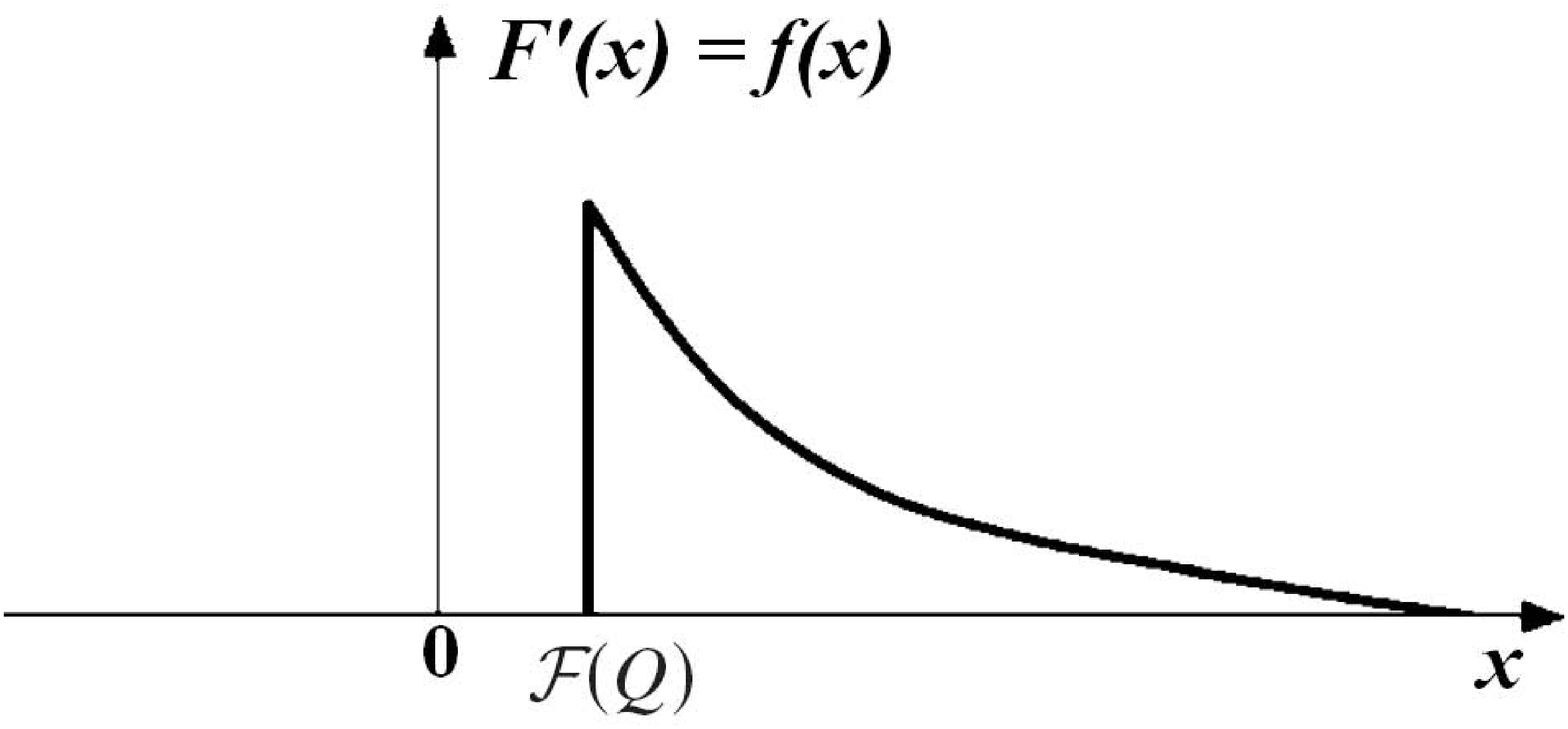, width=7cm, height =3cm}
     \caption{Density of the lead time profile $f(x)  = {dF(x) / dx}$ when ${\cal F}(Q) >0$.}
     \protect\label{Fig_FDEQPOS}
 \end{figure}

\begin{figure}[ht]
 \epsfig{file=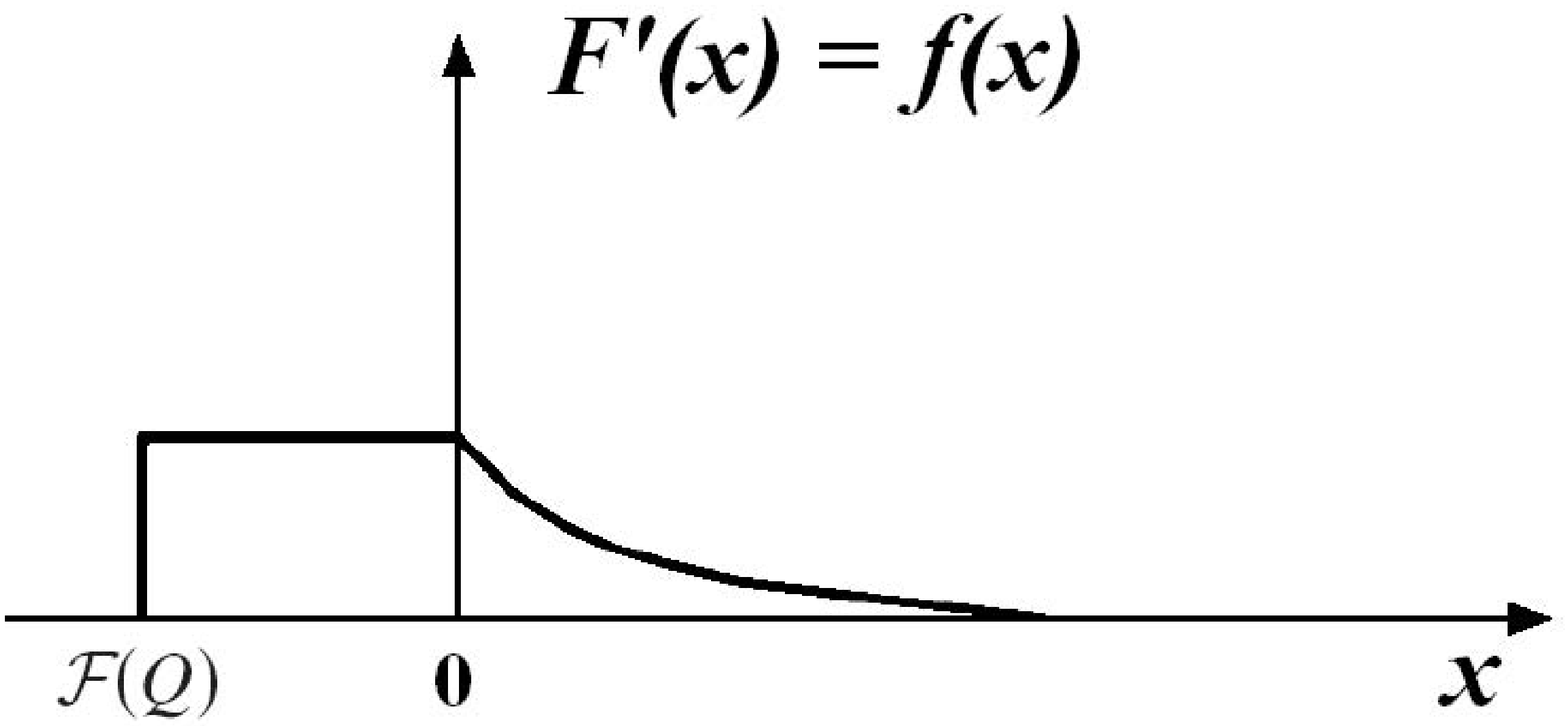, width=7cm, height =3cm}
     \caption{Density of the lead time profile $f(x) = {dF(x) / dx}$ when ${\cal F}(Q) <0.$}
     \protect\label{Fig_FDEQNEG}
 \end{figure}

 In \cite{Doytchinov:2001}, it is shown that two alternative regimes can occur:
\begin{itemize}
    \item[] a) {\it Jobs served before deadline}.
When $\langle \mathcal{D}\rangle>Q/\lambda $
 and therefore
$ \hat{\mathcal{F}}(Q)=\mathcal{F}(Q),$
the LTP cumulative distribution $F(x)$ takes
the form (see Fig.~\ref{Fig_FDEQPOS})

\begin{equation}\label{LTP1}
F(x) \,\,=\,\, \left\{
\begin{array}{l@{\qquad}l} 0 & {\rm when}\ x < {\cal F}(Q),\\ \, \\
1-\frac {\lambda}{\,\, Q\,\,}\left(\int_{x}^{\infty}\left[1-G(\eta)\right]
d\eta \right)& {\rm when}\ 0<{\cal F}(Q) \leq x.
\\
\end{array}
    \right.
\end{equation}

    \item[] b) {\it Jobs served after  deadline}.
When
$\mathcal{F}(Q)=\left(\langle\mathcal{D}\rangle-Q/\lambda
\right)\leq 0,$
the LTP
   cumulative distribution $F(x)$ takes the form
(Fig.~\ref{Fig_FDEQNEG})

\begin{equation}\label{LTP2}
F(x) =
\left\{
\begin{array}{lrr}
0,& {\rm when}\ x<  \left\langle{\cal D} \right\rangle -\frac{Q}{ \lambda}
<0,\\
\left[1+ \frac{\lambda}{Q}\left( x-\langle\mathcal{D}\rangle\right)\right]\,
\frac{Q-\lambda\langle\mathcal{D}\rangle}{Q-\lambda\langle\mathcal{D}\rangle+\lambda l},
& {\rm when} \
\left\langle{\cal D} \right\rangle -\frac{Q}{ \lambda} \leq x <l,\\
 1-  \frac {\lambda}{\,\, Q\,\,}\left\{\int_{x}^{\infty}\left[1-G(\eta)\right]d\eta\right\},& {\rm when} \ l \leq
 x.
\end{array}
\right.
\end{equation}

\end{itemize}

 \noindent {\bf Remark}. The alternative regimes given by (\ref{LTP1}) and (\ref{LTP2})
  can be heuristically understood by invoking the 
 {\it Little law} which connects the average
queue length $\langle Q \rangle$ with the average waiting time
$\langle W \rangle$, \cite{Cohen:1973},

\begin{equation}\label{LITTLE}
\langle Q \rangle\,\, =\,\, \lambda \, \langle W \rangle,
\end{equation}
a result independent of the scheduling policy. In view
of (\ref{FRONTIER}) and (\ref{LITTLE}), one obviously suspects
that the $LTP$  strongly
 depends on the sign of the difference
 $$
\langle{\cal
 D}\rangle-\frac{\langle Q \rangle}{\lambda}\,\,=\,\, \langle{\cal D}\rangle- \langle W\rangle.
$$
Intuitively, when $\langle W\rangle$ exceeds $\langle D \rangle$,
it is
 expected, in the average, that processed jobs will be delivered too late and conversely.
 While the above heuristic arguments is strictly valid only for the averages,
\cite{Doytchinov:2001},  were able to show that in heavy traffic regimes, it also holds also for the LTP
 given in  (\ref{LTP1}) and (\ref{LTP2}).

 Assuming that the arriving tasks have
positive deadlines, the LTP as given by  (\ref{LTP1}) and
(\ref{LTP2}) imply:

\begin{itemize}
    \item[] a). If the left-hand support of the LTP is negative,
    then a job entering into service
 is already late, (case of (\ref{LTP2})) see Fig.~\ref{Fig_FDEQNEG}.

\vspace{0.3cm}

    \item[] b). If the left-hand support of the LTP is positive
    then a   jobs enters into service with
 a positive lead time, (case of  (\ref{LTP1})) 
 see Fig.~\ref{Fig_FDEQPOS}.
Accordingly, it is likely that the tasks will
    be completed before the deadline expired.

\vspace{0.3cm}

    \item[] c). The critical value
$Q^*={\left\langle{\cal D} \right\rangle}/{\lambda}$  for which $\mathcal{F}(Q)=0$, corresponds to a queue length for which customers
    are likely to become  late. Choosing $Q$ 
exactly to $Q^{*}$, we cannot expect lateness to
    disappear completely but for $ Q < Q^{*}$ lateness will be strongly
    reduced a behavior clearly confirmed by  simulation
    experiments \cite{Baldwin:2000,Doytchinov:2001}.

\vspace{0.3cm}

 \item[] d).  For deadline distributions $G(x)$ with fat tails, it
 follows immediately from  (\ref{LTP1}) and (\ref{LTP2}) that
 the LTP does possess a fat tail.

\end{itemize}

\subsubsection{"First come  first served" (FCFS)  scheduling  policies}
    \noindent

Choosing the deadline probability density as
 $g(x) = \delta(x)$,
(i.e. zero deadline), the EDF scheduling policy directly coincides
 with  the FCFS rule. For this case we have $Q^*=0,$ and
(\ref{new_frontier0}) implies
\begin{equation}
\label{eq16}
\mathcal{F}(Q)\,\,=\,\,\left\{
\begin{array}{lrr}
\hat{\mathcal{F}}(Q)\,=\,0,& \mathrm{when} & Q\leq 0, \\
-\frac{\,\,Q\,\,}{\,\,\lambda\,\,}, & \mathrm{when} & Q> 0.
\end{array}
\right.
\end{equation}
Hence
the LTP
 density is given by  (\ref{LTP1})  is  merely the
uniform probability density
$U\left[- {Q/ \lambda},0\right]$,
 ($\left[- {Q/ \lambda},0\right]$ being its support).
 This
 expresses the fact that in the heavy traffic
 regime $\rho = \lambda / \mu \approx 1$, the waiting
 time behaves as $Q \times ({1/ \mu}) \approx Q \times ({1/
 \lambda})$  leading to a LPT linearly growing
 with $Q$. For  general $G(x)$, the LTP associated
 with a FCFS scheduling rule will
  be  given by the convolution of the deadline distribution $G(x)$
 with the uniform distribution $U\left[- {Q/\lambda},0\right]$.
 Indeed, adding the task deadlines with the
 time spent in the queue,  we recover the tasks lead-time.
 Accordingly, in the heavy traffic regime and for a given queue
length $Q$, one  explicitly knows the LTP's for both the EDF and
the FCFS scheduling policies thus enabling to explicitly
appreciate
 their respective characteristics. In particular, using
 (\ref{LTP1}) and (\ref{LTP2}), one can conclude that for a
given queue length $Q$, with the FCFS scheduling rule and the associated
LTP $F(x)$ being the convolution of $G(x)$ with the $U\left[-
{Q/\lambda},0\right]$, we obtain

\begin{equation}\label{CONVOLO}
    F(x) = \left\{
\begin{array}{l@{\qquad}l}
0 & {\rm when}\  x < -{Q/ \lambda},\\
\,
\\ \frac {\lambda}{Q}\int_{-\left({Q / \lambda}\right)}^{x}\left[G( \xi + {Q /
\lambda})\right]d\xi &{\rm when} -{Q / \lambda}\leq x < 0, \\
\,
\\  \kappa  +\frac{\lambda}{Q}
\left\{\int_{0}^{x}\left[G( \xi + {Q / \lambda}) - G(\xi)\right] d\xi \right\}&{\rm when}\ x\geq 0,\\
\end{array}
    \right.
\end{equation}

 where the constant $\kappa$ reads as:
$$
\kappa \,\,=\,\, \frac{\lambda}{Q}\int_{-\left({Q / \lambda}\right)}^{0}\left[G(\xi + \frac{Q}{\lambda})\right]d\xi.
 $$
The latter equation  allows to emphasize the following
features:

\begin{itemize}
    \item[] i). When the left-hand support of the deadline
 distribution $G(x)$ is
    larger than ${Q / \lambda}$, the left boundary of the
 support of $F(x)$ is
    larger than $0$ and therefore the jobs experience no delay
    when entering into service.

\vspace{0.3cm}

    \item[] ii). If the left-hand  support of $G(x)$ is
    smaller than ${Q / \lambda}$, then it may happen that
     the LTP exhibits a negative left-hand support under the FCFS
     policy and a positive left-hand support under the  EDF
    scheduling rule. Hence in this last situation, the FCFS
 policy would deliver tasks with lateness
    while the EDF  tasks will be processed in due time.
     This explicitly confirms intuition that  EDF is indeed an efficient
    policy. It has been shown that the  EDF
scheduling rule is optimal for minimizing the number of jobs
processed after the deadline \cite{Panwar:1988}.

\vspace{0.3cm}

    \item[] iii). If $G(x)$ exhibits a fat tail for $x \rightarrow \infty$ so has the
    LTP and this whatever the scheduling rule in use. This can e
    directly verified from  (\ref{CONVOLO}) by studying the LTP
    density $f(x) = {d F(x) / dx} $ for $x \rightarrow
    \infty$, we have:

    $$
    f(x) = \frac{\lambda}{Q}\left[ G(x+ \frac{Q}{\lambda}) - G(x) \right], \qquad \qquad
    {\rm for } \qquad  x \rightarrow \infty,
$$
\noindent which when  $G(x) \sim 1- x^{-q}$ and for ${Q /
\lambda} < {\rm const}$ takes the form

\begin{equation}\label{LTP-DENSSS}
f(x) \sim x^{-(q+1)}, \qquad \qquad
    {\rm for } \qquad  x  \rightarrow \infty.
\end{equation}

    Hence, the LTP  inherits the fat tail property of
    $G(x)$ and this even when using the optimal EDF scheduling
    rule.

\end{itemize}

Below, we focus on a fully explicit illustration involving the Pareto probability distribution
    \begin{equation}\label{PARETO}
    G(x)\,\, =\,\, \left\{
\begin{array}{l@{\qquad}l} 0 & {\rm when}\ {x < B},\\ \,
\\ 1- \left(\frac{B}{x}\right)^{(\omega-1)}
&{\rm when}\ \frac{x}{B} \geq 1, \qquad \omega > 1, \\
\end{array}
    \right.
\end{equation}
 which has  no moment of order $\omega-1$ or higher. For
$\omega >2$,  we have
$$
\left\langle{\cal D}\right\rangle\,\, =\,\,\left[\frac{
\omega-1}{\omega-2}\right]\, B.
$$
Using (\ref{new_frontier0}) with $l=B$, which
implies
$$
Q^*\,\,=\,\, \frac{\lambda B}{\,\,\omega-2\,\,},
$$
we obtain
\begin{equation}\label{FRONTIER-1}
    \mathcal{F}(Q) =
\left\{
\begin{array}{l@{\qquad}l}
B \left(\frac{B \lambda}{ Q(\omega - 2) }\right)^{1/(
\omega-2)}, & {\rm when}\ {Q } \leq  \frac{\lambda B}{\omega -2},\\
 \left[\frac{\omega-1}{\omega-2}\right]B- \frac{Q}{\lambda},
&{\rm when}\ {Q } > \frac{\lambda B}{\omega -2}. \\
\end{array}
    \right.
\end{equation}

 Using (\ref{LTP1}) and (\ref{LTP2}), we can find that the LTP
 distribution reads as

\begin{equation}\label{PAR1}
 Q  \geq \frac{B\lambda}{\omega -2} \quad  \Rightarrow \quad F(x)\,\, =\,\, \left\{
\begin{array}{l@{\qquad}l}
 0, & {\rm when}\ x \leq {\cal F} (Q),\\ \,
\\ 1 - \frac{\lambda}{ Q}\left(\frac{\omega -1}{\omega-2}\,B
-x\right),
&{\rm when}\ {\cal F} (Q) \leq x < B,  \\ \, \\
1- \frac{B \lambda}{ Q(\omega-2)} \left(\frac{B}{x}\right)^{\omega-2}.& {\rm when}\ x \geq B.
\end{array}
    \right.
\end{equation}
\begin{equation}\label{PAR2}
Q   < \frac{\lambda B}{\omega -2} \quad  \Rightarrow \quad
F(x)\,\, =\,\,\left\{
\begin{array}{l@{\qquad}l} 0, & {\rm when}\ x \leq {\cal F} (Q), \\ \,
\\ 1- \frac{B \lambda}{ Q(\omega-2)}\left(\frac{B}{x}\right)^{\omega
-2}.
&{\rm when}\ x > {\cal F} (Q). \\
\end{array}
    \right.
\end{equation}

The latter equations describe a  fat tail
with the exponent  $\omega-2$. It is worth to mention
that (\ref{PAR2}) implies
 that for ${\omega >2 }$ and for
$$
\frac{Q}{\lambda}\,\, <\,\, \frac{B}{\omega -2},
$$
the EDF scheduling policy  part of the  tasks enter into the service
before the due date expired. Finally note also,  that for $\omega
\leq 2$, no moments exists for the deadline distribution and hence
the theory \cite{Doytchinov:2001} cannot be applied directly. We
conjecture that for these regimes no scheduling rule will be able
to deliver tasks in due time.

 The results obtained for the LTP, enable us to investigate the
asymptotic properties of the  waiting time distribution.
Indeed, assume a heavy traffic regime with the
EDF scheduling policy.
Let us also suppose that for   a given queue
length, some jobs are served too late (i.e., the left boundary
of the LTP is negative). As under the EDF rule, the more
urgent jobs are always served first, the waiting times of the
last jobs in the queue necessarily exceed their deadlines.
Therefore, when the deadline distribution exhibits a fat tail,
so will the WTD distribution. Note that while the EDF policy
decreases, compared with the FCFS rule, the number of jobs
served after their deadline, it cannot get rid of the fat tail of
the WTD, which is due to the fat tail of $G(x)$. This result is
fundamentally different from the situation that is valid for
the frozen in time PI models discussed in \cite{Barabasi:2005,Vazquez:2005,Vazquez:2006}, where the
 fat tail behavior does not depend on $G(x)$ itself. This can be
heuristically understood as,
 in \cite{Barabasi:2005,Vazquez:2005,Vazquez:2006},
 the fat tail is mainly
due to the low priority jobs, which, as no aging mechanism
occurs, are likely to never be served. Note that in \cite{Barabasi:2005,Vazquez:2005,Vazquez:2006},
stable queuing models (i.e., those for which the traffic)
\cite{Barabasi:2005} and fat tails of the WTD disappear under
 a FCFS scheduling
rule. Indeed without priority scheduling, the WTD always
follows an exponential asymptotic decaying behavior.
In the presence of time-dependent PI, all tasks do finally
acquire a high priority and this aging mechanism precludes
the formation of a fat tail solely due to the scheduling rule.
Accordingly, in the presence of aging PI, the appearance of
WTD with fat tails is due to $G(x)$.

\subsection{The importance of adopting performance
scheduling policies}
\label{subsec:FAT_TAILS_3_3}

The results for the LTP derived in the preceding section
can be directly measured on the actual QS.
Consider the queue content of a single-stage QS. Assume
that at a given time, $Q$ is the observed queue content, and at
this instant take a snapshot of the lead time associated with
each waiting item and construct the associated LTP (i.e., the
histogram of the observed lead times). In heavy traffic regimes
(i.e., typically $0.95\leq\rho<1$ leading to stationary average
queue lengths $\rho/(1-\rho)$)
 and under the EDF scheduling policy,
the LTP will approximately be given by (\ref{LTP1},\ref{LTP2}).
Actual simulation experiments are reported in \cite{Lehozcky:1996,Lehozcky:1997,Doytchinov:2001}  and
\cite{Baldwin:2000}, where an excellent agreement between measured data
and theory is observed.

From the human activity viewpoint, the explicit expressions
of the LTP obtained both for the FIFO and EDS policies
show clearly that organizing the work scheduling is extremely
important. As an illustration, consider a situation in
which the deadline distribution $G(x)$ follows an exponential
law:
\begin{equation}
\label{deadline_distr}
G(x)\,\,=\,\,1-e^{-\alpha x}\,\,\Rightarrow\,\, \langle
\mathcal{D}\rangle =\frac 1\alpha.
\end{equation}
For this situation, we compare two different organization
policies:
\begin{enumerate}
\item {\it EDF scheduling policy.}\index{EDF scheduling policy}
 Introducing (\ref{deadline_distr}) into (\ref{FRONTIER}),
 we obtain
\begin{equation}
\label{EDF_SP}
\mathcal{F}(Q)\,\,=\,\,\left\{
\begin{array}{lrr}
-\frac{1}{\alpha}\log\left(\frac{\alpha Q}{\lambda}\right) & \mathrm{when} & Q\leq \frac \lambda\alpha, \\
\frac 1\alpha -\frac Q\lambda & \mathrm{when} & Q> \frac \lambda\alpha.
\end{array}
\right.
\end{equation}
When $\mathcal{F}(Q)>0$ (i.e., the upper line in  (\ref{EDF_SP})),
 it follows from  (\ref{LTP1}) that
\begin{equation}
\label{bozhe_moi}
F(x)\,\,=\,\,\left\{
\begin{array}{lrr}
0 & \mathrm{when} & x<\mathcal{F}(Q),\\
1-\frac{\lambda (1-\alpha x)}{\alpha Q} & \mathrm{when} & \mathcal{F}(Q)\leq x<0,\\
1-\frac{\lambda}{\alpha Q}e^{-\alpha x} & \mathrm{when} & 0\leq x.
\end{array}
\right.
\end{equation}
\item {\it FIFO scheduling policy.}

  With $G(x)$ given by (\ref{deadline_distr}),
the result given in (\ref{CONVOLO}) reads as
\begin{equation}
\label{nadoelo1}
F(x)\,=\,
\left\{
\begin{array}{lr}
0,& x< -\frac Q\lambda,\\
1+\frac{\lambda x}Q-\frac \lambda{\alpha Q}\left(1-e^{-\alpha\left(x+\frac Q\lambda\right)}\right),& -\frac Q\lambda\leq x<0,\\
1-\frac{\lambda\left(1-e^{-\frac{\alpha Q}{\lambda}}\right)}{\alpha Q}
e^{-\alpha x}, & x\geq 0.
\end{array}
\right.
\end{equation}

\end{enumerate}

Comparing (\ref{bozhe_moi}) and (\ref{nadoelo1}),
 we conclude that in a heavy
traffic regime, for a given work load $Q$, the use of EDF
enables us to process tasks in due time with a high probability
while the naive FIFO policy generates large delays.

 Specifically,
when $Q<\lambda/\alpha$, the EDF policy guarantees that most
jobs enter into service before the deadline (see (\ref{bozhe_moi})) and
will therefore be served before deadline, with a high probability.
On the contrary, the FIFO policy result given in (\ref{nadoelo1})
(i.e., obtained for $x=0$ in the last line of (\ref{nadoelo1}) shows
that a proportion of 
$1-\left(\lambda\left[1-e^{-\alpha Q/\lambda}\right]/\alpha Q\right)$
jobs enter the service
with delays and will therefore be late.

As far as human resources are concerned, this simple model enables
us to quantify the importance of adopting performance scheduling
policies to respond to the burn out -- generating challenge: {\it
deliver more in less time with fewer resources}. Along the same
lines, one of the key rules to avoid burnout is to {\it learn to
say no} to new incoming tasks if the queue length exceeds a
threshold. In our modeling framework, the critical threshold does
depend closely on the level $Q^*$, above which lately served tasks
(and hence complaints) are unavoidable.

\section{Power law distributions in Self--Organized Criticality}
\label{sec:FAT_TAILS_4}
\noindent

In the last two decades, a meticulous attention has been drawn to
the phenomenon of {\it self--organized criticality}
(SOC),
a property of
dynamical systems which have a critical point as an attractor,
 \cite{Bak:1987,Bak:1996,Zhang:1989,Jensen:1998}.
  A notable feature of these
models submitted to a power law statistics  is that they have no
characteristic scales, similarly to the scale invariant systems
being in a critical state. However, unlike  systems tackled  by
the critical phenomena theory the critical state in the SOC
models seems to be an attractor of the dynamics and seems to be
achieved without any tuning of control parameters.
It is observed in 
slowly-driven non-equilibrium systems 
with extended degrees of freedom and a high level of nonlinearity. 
The general idea behind SOC models is very appealing. 
Consider for instance {\it Zhang's sandpile
model} 
 on $\mathbb{Z}^2,$ where each site has an 
energy variable which evolves in discrete time-steps according
to a simple "toppling" rule: If a variable exceeds a threshold 
value, the excess is distributed equally
among the neighbors. The neighboring sites may thus
 turn supercritical and the process continues until
the excess is "thrown overboard" at the system boundary.

 What makes this dynamical rule intriguing is
that if the toppling is initiated from a
 "highly excited" state, then the terminal state 
(i.e., the state where
the toppling stops) is not the most stable state,
 but one of many least-stable, stable states. Moreover,
the latter state is critical in the sense that further
 insertion of a small excess typically leads to further
large-scale events. Using the sandpile analogy, such 
events are referred to as avalanches.

From the very beginning, large theoretical efforts have been made
in order to understand a true relation between criticality and
self-organized criticality both exhibiting a power law
behavior \cite{Sornette:1995,Bagnoli:1997}.
In particular, the use of
various {\it renormalization group techniques} (RG) which proved their
exceptional efficiency in justifying scaling properties in the
critical phenomena theory \cite{Ma:1976} has been in the focus of many
studies devoted to the SOC
phenomena,
\cite{Pietronero:1994,Vespignani:1995,Corral:1997,IPVZ:1999,Volchenkov:2002,Doussal:2008,Volchenkov:2009}..This  still deserves a
thorough investigation as a potential candidate for the "SOC
phenomena theory".

In the critical phenomena theory, the RG method usually helps to
establish the long time and large scale asymptotic behavior in
infinite systems defined by the stochastic differential equations
with the Gaussian distributed external random force \cite{Foster:1977}
that models random boundary conditions. RG is an effective method
of studying self-similar scaling behavior in such systems. On the
contrary, the majority of models exhibiting SOC phenomena are
defined on a finite piece $\mathcal{L}$ of a discrete lattice
$\mathbb{Z}^d$ by  discrete time dynamical
rules \cite{Bak:1987,Zhang:1989}. Moreover, in SOC models
the energy is usually dissipated at the open boundaries of the
lattice piece, while, in the majority of critical phenomena, a
quenched distribution of absorbing defects through the lattice is
imposed.

The primal goal for the SOC phenomena theory is to investigate the
scaling properties of  SOC models, in particularly, to justify the
numerically observed results \cite{Kadanoff:1989,Grassberger:1990}
 on the power
law distribution of avalanche sizes in sandpile models introduced
by Bak, Tang and Wiesenfeld, \cite{Bak:1987}. In more general
formulation, the finite size scaling (FSS) hypothesis \cite{Cardy:1984,Cardy:1988}
is usually assumed in SOC systems,
\begin{equation}\label{FSS}
P(x,L)=x^{-\tau_x}\mathcal{F}\left(\frac{x}{L^{\sigma_x}}\right),
\quad x\equiv \{s,{\ }t,{\ }a\},
\end{equation}
where $P(x,L)$ is the probability distribution of occurrence of
an avalanche of a given size $s$ (the number of sites involved in
a relaxation process), area $a,$ and time $t$,  ${\ }L$ is the
size of the lattice piece. If the FSS Ansatz (\ref{FSS}) is
valid, then the dynamical exponents $\tau_x$ and $\sigma_x$
determine the universality class of the model
\cite{Bak:1987}-\cite{Jensen:1998}. Within the framework of numerous
models, the dynamical exponents $\tau_x$ and $\sigma_x$ are found
from the different phenomenological approaches, which are loosely
related to underlying microscopic models, and therefore some
doubt remains about the universality of representations such as
(\ref{FSS}), \cite{Vespignani:2000,deMenech:2002,Tebaldi:1999}.

To identify the  scale invariant dynamics, the real-space RG
method had been applied to the cellular automaton defined on a
2-dimensional lattice\cite{Pietronero:1994,Vespignani:1995}.
 This approach ({\it Dynamically
Driven Renormalization Group} (DDRG)) deals with the critical
properties of the system by introducing in the renormalization
equations a dynamical steady state condition which assumes
non-equilibrium stationary statistical weights to be used in the
calculation \cite{IPVZ:1999}. The fixed points of scaling
transformations define the dynamical exponents whose value are in
a good agreement with computer simulation data. Nevertheless, it
has been shown \cite{Pietronero:1994} that the fixed points related to
these dynamical exponents prescribed by the renormalization group
are not accessible form the physical domain of parameter values.

An alternative approach referring to the standard critical
phenomena theory is based on the coarse-graining of microscopic
evolution rules for the SOC models. In \cite{Hwa:1989,Grinstein:1990} a continuous
stochastic partial differential equation related to the randomly
driven models had been proposed  although the threshold nature of
the SOC phenomena was not taken into account. A stochastic
partial differential equation subjected to a threshold condition
and driven by a Gaussian distributed external random force acting
continuously in time has been discussed in \cite{Corral:1997,DiazGuilera:1994,Perez:1996}.

Therein the external random force introduced into the dynamical
equation simultaneously models: first, the noise risen due to
elimination of microscopic degrees of freedom; second, unknown (or
undefined) boundary conditions; third, a mechanism  injecting
energy into the system which is supposed to act continuously in
time, breaking the time scale separation, and could provoke
avalanches to overlap. The threshold condition is taken into
account by the Heaviside step function $\theta(x)$. This step
function had been regularized as a limit of continuous infinitely
differentiable  functions and then expanded into power series
giving rise to an infinite series of nonlinearities in the
stochastic differential equation \cite{Corral:1997}.

Solutions of this nonlinear partial differential equation could
be found by iterating in the nonlinearities followed by averaging
over the distribution of the random force. Then, the long time
large scale asymptotic behavior of the solutions could be
established by means of a dynamic RG procedure in the spirit of
the dynamic RG-approach \cite{Foster:1977}. Particularly, the values of
dynamical exponents could be found in the form of power series in
$\varepsilon=4-d$. However, due to an infinite number of
nonlinearities risen in the stochastic differential equation by
the power expansion of step function, the resulting theory calls
for an infinite number of charges (coupling constants) and,
therefore, cannot be analyzed in the framework of the standard
dynamic RG scheme. In   \cite{Corral:1997}, only the first two
nonlinear terms have been kept for the RG analysis of the
appropriate stochastic problem. All higher order terms had been
neglected without any estimation of their contributions to the
long time large scale asymptotic behavior. Let us note that the
standard power counting analysis of such a model  shows
convincingly  that all these terms are of equal importance for the
asymptotic behavior and all of them have to be taken into
consideration on equal footing.

It is important to emphasize that the correspondence between the
models of  deterministic dynamics and the above stochastic
problem is indeed questionable and usually lays beyond the
studies. The obvious advantage of such a coarse graining approach
is that it allows to use the modern critical phenomena techniques
of analysis achieving impressive results on the self-similar
scaling behavior.

Here, we
present the results \cite{Volchenkov:2002,Volchenkov:2009}
on the long time large
scale asymptotic behavior for the model based on the nonlinear
stochastic dynamics equation derived from the coarse-graining
procedure from the discrete rules of deterministic dynamics
holding all nonlinear terms in check. This task is highly
nontrivial and of sufficient interest itself stimulating further
developments in  modern critical phenomena theory \cite{Ma:1976}.

\subsection{Coarse-graining of  microscopic evolution
 rules for SOC--models}
\label{subsec:FAT_TAILS_4_1}

Recently, two  randomly driven  SOC models proposed by
Zhang \cite{Zhang:1989} and by Bak \textit{et al.} 
\cite{Bak:1987} (BTW)  have
been connected to stochastic differential equations \cite{Corral:1997}.
For the convenience of the reader, we briefly describe the
microscopic rules of these SOC models. Both models are defined on
a finite piece of $d$-dimensional lattice
$\mathcal{L}\subset\mathbb{Z}^d$ in which any site
$i\in\mathcal{L}$ can store some continuously distributed variable
$E_i$ usually called energy \cite{Pietronero:1991}.

For Zhang's model, the system is perturbed by a random amount of
energy $\delta E>0$ at a randomly chosen site $i\in\mathcal{L}$.
Once the value $E_i$ exceeds a given threshold value $E_c$, this
site becomes active, and transfers all energy to the nearest
neighbors. As a result, the neighboring sites can be also
activated and transfer energy to the next neighbors, etc. until
it is absorbed at the open boundary $\partial \mathcal{L}$. The
avalanche  ends when all sites reached a value of energy smaller
than $E_c.$ The next energy input into the system occurs only
when the avalanche has stopped.

For the BTW model, the amount of energy  perturbing the system is
fixed $\delta E=E_c/q$ where $q$ is a coordination number, and the
amount of energy transferred to neighbors from an active site is
also fixed at $E_c$.

For both models, energy is pumped into the system at the
small-scale of lattice spacing $a$ and then is transferred to a
large scale comparable to the size of $\mathcal{L}$ and actively
dissipated at the open boundaries.

For each $i \in \mathcal{L}$, the microscopic evolution rules  can
be written in the form of a {\it stochastic coupled map
 lattice}\index{stochastic coupled map lattice} (SCML),
\begin{equation}\label{micro}
  E_i(t+1)-E_i(t)
  \hspace{8cm}
\end{equation}
$$
=\frac 1q \sum_{mm}
  \left[(\chi E_{mm}(t)+E_c)\theta\left(E_{mm}(t)\right)
  -(\chi E_i(t)+E_c)\theta\left(E_{i}(t)\right)
  \right]+\zeta_i(E,t),
$$
in which $E_i(t)$ is the exceed of energy over the critical value
$E_c$,  ${\ }\chi=1$ for Zhang's model and $\chi=0$ for BTW. The
external noise
\begin{equation}\label{noise}
  \zeta_i(E,t)=(\delta E)\cdot \delta_{i,\mathbf{n}(t)}\prod_{j\in \mathcal{L}}
  \left[
1-\theta\left( E_j(t)\right)
  \right]
\end{equation}
acts at a slow time scale, and is present when there are no
active sites in the lattice. Here $\mathbf{n}(t)$ is a random
vector pointing the site of the lattice piece $\mathcal{L}$ that
is perturbed with energy $(\delta E)>0.$

The dynamics of avalanches governed by the SCML (\ref{micro})
evolves infinitely fast in comparison with the dynamics of energy
feeding. It has been pointed out \cite{Corral:1997} that the SCML
(\ref{micro}) is invariant under spatial translations, rotations,
and reflections. Furthermore, if $\chi=0,$ the equation is
invariant under a parity transformation of the order parameter,
$E\to -E.$ The SCML (\ref{micro}-\ref{noise}) is
supplied with the absorbing boundary condition $E_{\partial
\mathcal{L}}(t)=0.$

The SCML (\ref{micro}) can be coarse-grained in order to
obtain a continuum stochastic differential equation \cite{Corral:1997} for
the effective continuum scalar field $E(\mathbf{r},t)$,
\begin{equation}\label{Eq1}
\frac{\partial E(\mathbf{r},t)}{\partial t}=
\alpha_0 \Delta \left[\left(\chi E(\mathbf{r},t)+
E_c\right)\theta\left(E(\mathbf{r},t)\right)\right]+
f(\mathbf{r},t)
\end{equation}
where $\alpha_0>0$ is the only dimensional parameter in the model,
$[\alpha]=L^2T^{-1},$ and depends on the lattice spacing $a$, the
unit time step, and the coordination number $q$. $\Delta$ is the
Laplace operator. The noise $f(\mathbf{r},t)$ is a sum of the
multiplicative external noise depending on the whole lattice
state and the internal noise that appears due to the elimination
of microscopic degrees of freedom. In the continuous model,
energy is thought to disappear in those regions of lattice where
$f(\mathbf{r},t)<0$ and to arrive at the points for which
$f(\mathbf{r},t)>0.$ In a stationary state, these processes  are
obviously balanced, therefore,  $\left\langle
f(\mathbf{r},t)\right\rangle=0.$

In  \cite{Corral:1997}, the important effect of dissipation at
the open boundaries has not been taken into account and a
quenched distribution of energy absorbing defects has been
assumed. Time scale separation of dynamics has been also
neglected, and the noise $f(x),$ ${\ } x\equiv \mathbf{r},t,$ has
been understood just as a quenched Gaussian process uncorrelated
in space and time with a covariance
\begin{equation}\label{rw}
  \left\langle f(x)f(x') \right\rangle=
  2\Gamma \delta^d(\mathbf{r}-\mathbf{r'})\delta(t-t'),
\end{equation}
which is typical for random walks. In the present section, we use
a different Ansatz for the covariance $\left\langle f(x)f(x')
\right\rangle$ which takes  the slow-time scale dynamics of the
stochastic force $f(x)$ into account (see the next section).

The continuum stochastic partial differential equation
(\ref{Eq1}) requires a regularization of the step function at
zero. Following \cite{Corral:1997}, we use
\begin{equation}\label{reg}
  \theta(E)=\lim _{\Omega\to\infty} \frac {1+\mathrm{erf}(\Omega
  E)}2,
\end{equation}
as a regularizing function   where  $\mathrm{erf}(x)=
\pi^{-1/2}\int^x_{-\infty}\exp [-y^2]{\ }dy$ is the error
function and $\Omega$ is the regularization parameter. The reason
for this choice of regularization procedure is that it allows a
power expansion  with an infinite radius of convergence.

Developing (\ref{reg}) in powers of $E$ and substituting the
series expansion into (\ref{Eq1}), one obtains the strongly
nonlinear stochastic partial differential equation
\begin{equation}\label{Eq2}
\frac{\partial E(\mathbf{r},t)}{\partial t}=
\alpha_0 \sum_{n\geq 1}^{\infty}\frac{\lambda_n{}_0}{n!}
\Delta E^n(\mathbf{r},t)+ f(\mathbf{r},t),
\end{equation}
where the effective coupling constants take different values
depending on the model:
\begin{equation}\label{lam}
  \lambda_n{}_0=\lim_{\epsilon\to \infty} \epsilon^n\left[E_c \theta^{(n)}(0)+
  \frac{n\chi}{\epsilon}\theta^{(n-1)}(0)\right], \quad n\in \mathbb{N}
\end{equation}
in which $\theta^{(n)}(0)$ is the $n-$th order derivative of the
regularizing function (\ref{reg}) at zero. The coefficient $
\lambda_n{}_0$ becomes formally infinite as $\epsilon\to\infty,$
however, the series in (\ref{Eq2}) converges. In the equation
(\ref{Eq2}), we have supplied the parameter $\alpha_0$ and  the
coupling constants $\lambda_n{}_0$ with index $"0"$ to
distinguish their bare values from the renormalized analogs which
we shall denote in forthcoming sections simply as $\alpha$ and
$\lambda_n$ consequently. It has been noted \cite{Corral:1997} that since
$\theta^{(2n+2)}(0)=0,$ all even coupling constants vanish for
the BTW model, whereas they do not for the Zhang's one. The set
of coupling constants $\lambda_n{}_0$ for both models are
identical in the limit $\epsilon\to \infty.$

The equation (\ref{Eq2}) describes the diffusion of energy in
$\mathbb{Z}^d$ issued from a source defined by $f(x),$ ${\
}x\equiv\mathbf{r},t$. This equation (up to a minor change of
notations) has been considered in the work \cite{Corral:1997}  in the
whole space and the important effect of dissipation at the open
boundaries has not been taken into account. Alternatively, a small
probability of dissipating an amount of energy $E_c/q$ has been
assigned for each site when it topples, instead of transferring
it to a certain neighbor. This procedure expresses the
 \textit{ assumption
of random boundaries} and corresponds to a model defined
on an infinite lattice with a dissipation for each toppling site.
We discuss the possible changes to the critical behavior due to
the presence of regular absorbing boundary in the section 11.

We study the long time large scale
asymptotic behavior in the system governed by the stochastic
differential equation (\ref{Eq2}) in the whole space supposing
that the random force $f(x)$ is Gaussian distributed and
characterized by the covariance (see the next section) which goes
beyond the "white noise" approximation studied in\cite{Corral:1997}. The
introduction of the random force $f(x)$ in (\ref{Eq2}) expresses
the boundary conditions at the \textit{ random
boundaries} \cite{Corral:1997}.

\subsection{Covariance of random forces}

The introduction of the random force into the equation (\ref{Eq2})
phenomenologically models  a consequence of the elimination of
microscopic degrees of freedom and, at the same time, the
injection of energy into the system. We take the time scale
separation into account supposing that the dynamics of the
slow-time scale and fast-time scale components of the random force
are essentially different. Namely, we suppose that in the
slow-time scale (i.e., the time scale of energy injection) this
dynamics can be taken as the white noise like in \cite{Corral:1997},
$\langle F(x)\rangle=0,$ with the covariance
\begin{equation}\label{rw010}
 D_F\equiv \left\langle F(x)F(x') \right\rangle=
  2\Gamma \delta^d(\mathbf{r}-\mathbf{r'})\delta(t-t'),
\end{equation}
where $\Gamma$ is the Onsager coefficient. However, in the
equation (\ref{Eq2}) defining the dynamics of relaxation processes
evolving in the fast-time scale \cite{Corral:1997}, the random force
$f(x)$introduced into r.h.s. has to be also of fast-time scale.
We take it as the generalized random walks governed by the linear
Langevin equation
\begin{equation}\label{Lsoc}
  \frac{\partial f(x)}{\partial t}+ \mathfrak{R}
  f(x)= F(x), \quad x\equiv \mathbf{r},t,
\end{equation}
driven by the slow-time scale "white noise" $F(x),$ where the
kernel of the pseudo-differential operator $\mathfrak{R}$   has
the form
\begin{equation}\label{kernel}
  \mathfrak{R}(k)=\rho_0 \alpha_0 k^{2-2\eta}
\end{equation}
in the Fourier space. Here, the coupling constant $\rho_0 >0$  and
the exponent $2\eta>0$ are related to the reciprocal correlation
time at wave number $k,$ ${\ }t_c(k)=k^{2\eta-2}/\rho_0\alpha_0.$
Dimensional considerations show that the coupling constant
$\rho_0$ is related to the characteristic  {\it ultra-violet} (UV) \index{ultra-violet momentum scale, UV}
 ultra-violet momentum scale in SOC
momentum scale $\Lambda\simeq 1/a$ by $\rho_0\simeq
\Lambda^{2\eta}$ and corresponds to microscopic degrees of freedom
expelled from the main equation (\ref{Eq2}) as a result of the
coarse-graining of deterministic dynamical rules \cite{Corral:1997}.

The exponent $\eta$ corresponds clearly to the anomalous diffusion
coefficient \cite{Jensen:1998,Cessac:2001} $z=2(1-\eta)$. Let us note that the
scaling form of   reciprocal correlation time $t_c(k)$ has
interesting connections with the spectrum  of Lyapunov exponent,
for the Zhang model. It has been indeed shown \cite{Cessac:2001} that the
Lyapunov exponents and the corresponding modes relate to  the
energy transport in the lattice. However, the transport in SOC
model is anomalous and the transport modes correspond to
diffusion modes in a non-flat  metric given by the probability
that a site is active. It is remarkable that the Lyapunov
spectrum obeys a simple finite scaling form, with an universal
exponent $\tau<1$, which is directly related to  $\eta$ by the
relation 
$$
\eta\,\,=\,\,\frac{d}{\,2(1-\tau)\,}.
$$
 The parameter $\rho_0$ corresponds
to the energy injection rate.

 It is believed in the literature
that the SOC regime corresponds to the case when the injection
rate goes to zero, the dissipation rate goes to zero, such that
the ratio injection/dissipation goes to zero establishing the
time scale separation \cite{Pietronero:1994,Vespignani:1995}.

In the framework of critical phenomena theory approach to the
problems of stochastic dynamics 
(see \cite{Zinn:1990, Adzhemyan:1998}, for example),
the model for the random force covariance $D_F$ in (\ref{rw010}) is
chosen under the following reasons:

i) for the use of the standard quantum-field RG technique, it is
important that the function $D_F$ have a power-law asymptote at
large $k$;

ii) since the covariance in (\ref{rw010}) is static ($\propto
\delta(t-t')$), the required \textit{physical} dimension
$\left[\langle FF\rangle\right]=L^d T^{-3}$ is put on by a
suitable combination of the only dimensional parameters in the
logarithmic theory ($\alpha_0$ and $k$);

iii) the "white noise" assumption (\ref{rw010}) means that
$D_F\propto\mathrm{Const}$ in the Fourier space. To meet this
requirement, one introduces a regularization parameter
$\varepsilon>0$ quantifying  the deviation form the logarithmic
behavior that is similar to the well-know $\varepsilon=4-d$
expansion parameter in the critical phenomena theory \cite{Ma:1976}.

All above requirements are satisfied by the model
\begin{equation}\label{DF}
\left\langle
F(\mathbf{k},\omega)F(-\mathbf{k},\omega')\right\rangle\equiv
D_F(k) \propto \alpha^3_0k^{6-d-2\varepsilon}.
\end{equation}
In this model, $2\varepsilon$  is completely unrelated to the
space dimension $d$ (in contrast to the standard critical
phenomena approach \cite{Ma:1976,Zinn:1990}, 
where usually $\varepsilon=4-d$).
The logarithmic theory
corresponds to the value $\varepsilon=0.$

Finally, the model for the covariance $D_F(k)$ has to be
consistent with the form of the linear operator (\ref{kernel}).
Namely, from the equation (\ref{Lsoc}), it follows  that the
covariance $D_f(\omega,k)$ for the pseudo-random force $f$
introduced in the r.h.s. of the main equation (\ref{Eq2}),
\begin{equation}\label{corr}
\left\langle f(x)f(x') \right\rangle=
\int \frac{d\omega}{2\pi} \int \frac{d\mathbf{k}}{(2\pi)^d}
D_f(\omega, k)\exp\left[ -i\omega
(t-t')+i\mathbf{k}(\mathbf{r}-\mathbf{r'})\right], \quad k\equiv
|\mathbf{k}|,
\end{equation}
is related to $D_F(k)$ as
\begin{equation}\label{Df}
D_f(\omega, k)=
\frac{D_F(k)}
{\omega^2+[\rho_0\alpha_0k^{2-2\eta}]^2}.
\end{equation}
Then, it is natural that the spectral density of energy injection,
\begin{equation}\label{Wk}
  \widetilde{W}(k)=\frac 12\int \frac {d\omega}{2\pi} D_f(\omega,k),
\end{equation}
is independent of the correlation time at given wave number,
$t_c(k)$. This is true if one takes $D_F(k)\propto \rho_0
k^{-2\eta}.$ Eventually, collecting the latter result with the
previous Ansatz (\ref{DF}), one arrives at the model
\begin{equation}\label{DF1}
  D_F(k)=\rho_0\alpha^3_0k^{6-d-2\varepsilon-2\eta}.
\end{equation}
Both exponents $2\eta$ and $2\varepsilon$ in (\ref{DF1}) are the
parameters of the double expansion in the $\eta-\varepsilon$ plane
around the origin $\eta=\varepsilon=0,$ with the additional
convention that $\varepsilon=O(\eta).$ The positive amplitude
factor $\rho_0k^{-2\eta}$ is considered as a dimensionless
coupling constant (i.e., a formal small parameter of the ordinary
perturbation theory). For the case of random force uncorrelated in
space, $D_F(k)\propto\mathrm{Const}$, the "real" values of
$\varepsilon$ and $\eta$ are taken such that
$6-d=2(\eta+\varepsilon).$ The similar power-law Ansatz for the
correlator of random force has been used to model the energy pump
into the inertial range of fully developed
turbulence \cite{Dominicis:1979,Adzhemyan:1998}.

The model (\ref{Df}) where the function $D_F(k)$ is defined by
(\ref{DF1}) is then more realistic and more reach in behavior than
the simple "white noise" assumption (\ref{rw}) discussed in the
literature before (for example in the work \cite{Corral:1997}) since it
takes into account the finite correlation time of energy field
set by interactions at a level of microscopic degrees of freedom.
It has a formal resemblance with the models of random walks in
random environment with long-range correlations. We note that the
similar correlator for random force has been discussed for the
first time in studies devoted to the anomalous scaling of a
passive scalar advected by the synthetic compressible
flow \cite{Antonov:1999}.

The Ansatz (\ref{Df}) that we use contains the previously
discussed \cite{Corral:1997} model (\ref{rw}) as a special case. Indeed,
for the rapid-change limit $\rho_0\to \infty,$ the covariance
(\ref{Df}) has the form
\begin{equation}\label{Isoc}
  D_f(\omega,k)\to \frac
  {\alpha_0}{\rho_0}k^{2-d-2\varepsilon+2\eta},
\end{equation}
and, for $\varepsilon-\eta=1-d/2$, one arrives at the model
(\ref{rw}) uncorrelated in space and time  with $\Gamma=
{\alpha_0}/{2\rho_0}$.

In the opposite limit of "frozen" configuration of the stochastic
force, $\rho_0\to 0$,  the covariance is static (i.e.,
independent of time argument $(t-t')$ in $t$-representation),
\begin{equation}\label{II}
D_f(\omega,k)\to  \pi\rho_0\alpha^2_0
k^{4-d-\varepsilon}\delta(\omega).
\end{equation}
The latter case obviously corresponds to an  external random
force acting continuously in time. For $\varepsilon=4-d$, this
random force is uncorrelated in space ($\propto
\delta(\mathbf{r}-\mathbf{r'})$).

\subsection{An infinite number of critical regimes in SOC-- models}

Quantum field theory formulation 
for SOC models has been introduced and studied in 
\cite{Volchenkov:2009} following the general approach 
developed in \cite{Wyld:1961, Martin:1973,Dominicis:1978,Dominicis:1979,Adzhemyan:1998}.
All correlation functions $\left\langle E(x_1)\ldots
E(x_k)\right\rangle, $ ${\ }x\equiv \mathbf{r},t,$ and response
functions $$\langle \delta^m \left[E(x_1)\ldots
E(x_k)\right]/\delta f(x'_1)\ldots \delta f(x'_m) \rangle$$
expressing the system  response for an external perturbation 
were renormalized by subtracting all ultra--violet
superficial divergences 
from  Green's functions.
 An infinite number of renormalization constants has
 been calculated in the one--loop approximation in \cite{Volchenkov:2009}.

Possible scaling regimes of a renormalizable model are associated
with the infra--red (IR) stable fixed points of the corresponding differential
RG equation. The coordinates of fixed points
$\{\rho_*,\lambda_n{}_*\}$ in the infinitely dimensional space of
coupling constants $\lambda_n$ of the stochastic model 
(\ref{Eq2})
 are the solutions of 
the equations 
\begin{equation}\label{beta0}
  \beta_\rho(\rho_*,\lambda_n{}_*
  )=\beta_n(\rho_*,\lambda_n{}_*)=0, \quad n=1,2,\ldots,\infty
\end{equation}
where $\beta_{\rho,n}-$functions, the coefficients in the 
differential RG equation, are some rational functions 
of the parameters $\rho$ and $\lambda_n.$
 Any solution of 
 (\ref{beta0}) is an IR-attractive (IR-stable) fixed point
of the RG equation  if the
corresponding Jacobian matrix 
$$
J_{ik}\,\,=\,\,\frac{\,\,\partial \beta_i\,\,}{\partial
\lambda_k}
$$ 
is positively defined (i.e., the real parts of all eigenvalues
of the matrix $J_{ik}$
are positive) for small $\eta>0,$ ${\ }\varepsilon>\eta$, $0<\rho<1$, where $\lambda_0\equiv \rho$ and $\beta_i$
denotes the complete set of the $\beta-$functions of the RG--equation.

It follows from the explicit expressions for an infinite set of $\beta-$functions 
found in \cite{Volchenkov:2009} 
 that two coordinates of
fixed points, $\lambda_1{}_*$ and $\lambda_2{}_*,$ can be chosen
arbitrary, then all other coordinates $\rho_*$ and
$\lambda_k{}_*$, ${\ }k>2$  can be found directly from the equations
(\ref{beta0}). 
Therefore, the RG differential equation 
for SOC models
 has a
two-dimensional surface of fixed points spanned with
$\lambda_1{}_*$ and $\lambda_2{}_*$ in the infinite dimensional
space of coupling constants $\{\rho, \lambda_n\}$.

The complete IR--stability analysis  for this manifold
of fixed points is a formidable task. 
In the limiting case of
"white noise" model (\ref{rw}), taking formally 
$$
\rho_*\,\,\to\,\,\infty,\quad \eta\,\,=\,\,0,
$$ 
it is possible to demonstrate that  
$$
J_{ik}\,\,\approx\,\,
-(n-1)\varepsilon \delta_{ik}\,\,<\,\,0,
$$ 
where $\delta_{ik}$ is the
Kronecker symbol, 
so that 
there are no IR-- stable fixed points in SOC
for  zero correlation time at all wave
numbers. 
The time scale
separation is mandatory 
for the existence of a critical regime in SOC.

In the opposite limit of "frozen" configuration of the stochastic
force,
the fixed points of the RG equation
have been shown also IR unstable, since
$$
J_{ik}\,\,=\,\,-(n-1)\varepsilon\delta_{ik}\,\,<\,\,0.
$$ 
In a general setting, the matrix $J_{ik}$ in such a case has  a
block triangular form, and therefore
 its eigenvalues coincide  with
the diagonal elements which can be calculated
for any $\beta_n.$
The stability domains are
defined by the roots of polynomials in $\rho_*$. 
For instance, 
the positivity of the first eigenvalues
requires that
$$
\frac{\rho_*^3+3\rho_*^2-2-4\rho_*}{(1+\rho_*)(1+\rho^2_*)}<0,
$$
and
$$
\frac{2\rho_*^2-3-5\rho_*}{(1+\rho_*)(1+\rho^2_*)}<0
$$
that is
true for $|\rho|<1.$
As $n$ grows up,
polynomials of any large order can appear splitting the stability
domain into a number of stable and unstable strips.

It is important to note that in a  multi-charge theory, even if
the IR-stable fixed points of the RG equation exist, the actual
trajectory of the system in the multi-dimensional (phase) space of
coupling constants (in our case, an infinitely dimensional space
$\{\rho, \lambda_k\}$) starting from the given initial values
$\rho_0, {\ }\lambda_k{}_0$ may not achieve any of them. The
trajectory can leave the stability domain (in the critical
phenomena theory, it is usually interpreted as the first order
phase transition \cite{Ma:1976,Zinn:1990}) breaking the scaling asymptote.

\subsection{Fat tails in SOC-models}

In the IR--stable critical regimes, the Green functions and the response functions exhibit scaling behavior characterized by the following "critical dimensions":
\begin{equation}\label{coeff-cs}
\begin{array}{l}
\Delta[t]\,\,=\,\,-\Delta[\omega]\,\,=\,\,-2+\gamma_\alpha{}_*\,\,=\,\,2\eta-2, \\
 \Delta[E]\,\,=\,\,2\eta-\varepsilon, \\
\Delta[E']\,\,=\,\,d+\varepsilon-2\eta.
\end{array}
\end{equation}
We have pointed out before 
that for the random force
uncorrelated in space ($D_F(k)\propto \mathrm{Const}$),
 the "real"
value $\varepsilon_r$ is taken such that
$$
3-d/2\,\,=\,\,\eta+\varepsilon_r.
$$ 
In two alternative limiting cases, we
have 
$$
\varepsilon_r\,\,=\,\,1-d/2+\eta
$$ 
(the "white noise" assumption; the
system lacks of IR-attractive fixed points) and
$\varepsilon_r=4-d$ (the "frozen" configuration of the random
force). Substituting these values into (\ref{coeff-cs}), we obtain
different  critical dimensions for the energy field $\Delta[E]$
and the auxiliary filed $\Delta[E']$ listed in the
Tab.~\ref{T2soc}.
\begin{table}[ht]
\caption{Critical dimensions of the fields $E$ and $E'$ depending on the various
 models for the covariance $D_F$ \label{T2soc}}
\centerline{\footnotesize  }
\centerline{\footnotesize
\begin{tabular}{c c c}
\hline \hline
$\varepsilon $& $\Delta[E]$ & $\Delta[E']$ \\
\hline
$3-\eta-d/2$ & $d/2+3(\eta-1)$ & $d/2+3(1-\eta)$\\
$1-d/2+\eta$ & $d/2+\eta-1$ & $d/2+1-\eta$  \\
$4-d$ & $2\eta-4+d$ & $4-2\eta$ \\
\hline \hline
\end{tabular}}
\end{table}
For instance,
for the critical dimension of 
the simplest Green function $<EE>(\omega,k),$
we obtain 
 $$
\Delta\left[<EE>
\right]\,\,=\,\,2\Delta[E]-d+\Delta[t],
$$
and
$$
\Delta[<EE>_{\mathrm{st}}]\,\,=\,\,2\Delta[E]-d,
$$
  for its static
analog, 
$$
<EE>_{\mathrm{st}}(k)\,\,=\,\,(2\pi)^{-1}\int
d\omega {\ }<EE>^R(\omega,k).
$$
 The simplest response function $<E'E>$  
evaluates the average size of the
relaxation process arisen in the system as a reaction for a
point-wise perturbation; its Fourier transform is the
distribution of avalanche size $P(s)$  observed in numerical
experiments. For $<E'E>,$ in the IR--stable critical regime
we obtain   
\begin{equation}
\begin{array}{lcl}
\Delta\left[\langle E'E\rangle\right]&=&\Delta[E']\\+\Delta[E]-d+\Delta[t]
&=&-2+2\eta.
\end{array}
\end{equation}
The squared  effective radius
\begin{equation}\label{sqefrad}
 R^2\,\,=\,\,\int d\mathbf{x}{\ }\mathbf{x}^2 \left\langle
E(\mathbf{x},t)E'(\mathbf{0},0)
 \right\rangle
 \end{equation}
of the relaxation process at a moment of time $t>0$ started at
$t'=0$ at the origin $\mathbf{x}=0,$ one can find that it scales
as 
\begin{equation}
\label{richardson01}
\frac{dR^2}{dt}\,\, \propto\,\, R^{2\eta}.
\end{equation}
Indeed, since $\Delta[R]=-1$
(by convention), 
$$
\Delta\left[\frac{dR^2}{dt}\right]\,\,=\,\, -2-\Delta[t],
$$ 
from where we
have the result (\ref{richardson01}). The obtained relation is
analogous to the well-known Richardson's phenomenological law for
the diffusion of passive admixtures in the ambient turbulent
flows \cite{Monin:1971}.

We have studied the long time large scale
asymptotic behavior for the strongly nonlinear stochastic problem
which relates both to the Zhang and BTW models exhibiting the
self-organized critical behavior. The proposed model is
interesting as itself since it is connected to the problem of
nonlinear diffusion of the chemically active scalar advection in
the turbulent flows \cite{Antonov:1997}. The stochastic problem  has
been considered in the bulk, far from a regular boundary. The
energy dissipation at the open boundaries has not been taken into
account, instead a quenched distribution of energy absorbing
defects has been assumed.

We now make several important comments on the further
investigations in the framework of RG approach to the stochastic
differential equations  related to the SOC
models.

The first comment is on the possible changes for the critical
behavior close to the regular open boundary exhibiting the
absorbing property. The equation (\ref{Eq2}) can be considered in
a half-space $z>0$ where the open boundary coincides with the
$z=0$ plane. Then the  effect of boundary would be due to the
semi-infinite geometry of the system: The absence of sites from
one-half space ($z<0$) changes the energy transfer along the
surface. Using the critical phenomena analogy \cite{Brezin:1982},
 one can
say that the surface does not become critical simultaneously with
the bulk, but tends to decouple from the rest of the system.
Furthermore, the pseudo-random force acting at the boundary is to
be always negative to ensure the complete dissipation of energy,
\begin{equation}\label{boundsf}
  h_\bot\equiv \left. f \right|_{z=0}(\mathbf{r},t)<0.
\end{equation}
This is equivalent to introducing the new field $h^0_\bot$ on the
surface that provokes a perturbation which can spread inside the
system. These two effects are in competition: If the coordination
number $q$ is large, the toppled amount of energy $E_c/q$
dissipated at the open boundary  is much smaller than that one
transferred to neighbors. In this case, the perturbation risen
due to $h_\bot$ close to the boundary cannot propagate into bulk.
Otherwise, if energy is rather intensively dissipated at the
boundary than transferred to the neighboring sites, a critical
slope can appear.

Another comment is on possible steps beyond the Gaussian
approximation. Let us note that the study of composite operators
of the type $(E')^n(x) $ would manage the corrections for the
non-Gaussian distributions of random force.  We also remember
that composite operators are important for the definition of finite
size scaling corrections  to the leading RG
predicted asymptotes.

Scaling, renormalization and statistical conservation laws in
 the Kraichnan model of turbulent advection in the context of the renormalization
group improved perturbation theory have been investigated
in \cite{Kupiainen:2007}.

\section{Conclusions }
\label{sec:FAT_TAILS_SUMMARY}
\noindent

Power laws (heavy-tailed) distributions
 are found throughout many naturally
 occurring phenomena in physics, and efforts to
 observe and validate them are an active area of
 scientific research.
We have considered 
a number of stochastic dynamical models 
 that might
 generate power law asymptotic distributions.
In particular, we have reviewed
 the stochastic processes
involving multiplicative noise,
Degree-Mass-Action principle (generalized 
preferential attachment principle),
 the intermittent behavior occurring in more
 complex physical systems near a bifurcation point, 
some cases of queuing systems, 
and 
the models of Self-organized criticality.

These models might be a ground for many 
natural complex systems.
Heavy-tailed  distributions
appear in them as the 
 emergent phenomena
sensitive for 
coupling rules essential for
 the entire 
dynamics. 
Relationships between 
the rules and the power law statistics 
are strikingly non-linear, as
even 
a small perturbation may cause 
a large effect,
 a proportional effect, or even no effect at all.

\section{Acknowledgments}
\label{sec:Acknowledgments}
\noindent

We would like to thank
Bruno Cessac,
Elena Floriani,
Max Hongler, and
Ricardo Lima for numerous discussions.

\end{document}